\documentclass[reprint, prb,
superscriptaddress,
amsmath,
amssymb,
aps
]{revtex4-2}

\usepackage{graphicx}
\usepackage{dcolumn}
\usepackage{bm}
\usepackage{hyperref}
\hypersetup{colorlinks=true,allcolors=blue}
\usepackage[utf8]{inputenc}

\usepackage{physics}
\usepackage{graphics}
\usepackage{xcolor}
\usepackage[caption=false]{subfig}
\usepackage{simpler-wick}
\usepackage{siunitx}

\usepackage{mathtools}

\DeclarePairedDelimiter\floor{\lfloor}{\rfloor}

\begin{document}

\title{Unique Signatures of Topological Phases in Two-Dimensional THz Spectroscopy}

\author{Felix Gerken}
\affiliation{I. Institut für Theoretische Physik, Universität Hamburg, Notkestraße 9, 22607 Hamburg, Germany}
\affiliation{The Hamburg Centre for Ultrafast Imaging, Luruper Chaussee 149, 22761 Hamburg, Germany}
\author{Thore Posske}
\affiliation{I. Institut für Theoretische Physik, Universität Hamburg, Notkestraße 9, 22607 Hamburg, Germany}
\affiliation{The Hamburg Centre for Ultrafast Imaging, Luruper Chaussee 149, 22761 Hamburg, Germany}
\author{Shaul Mukamel}
\affiliation{Departments of Chemistry and Physics $\&$ Astronomy, University of California, Irvine, California 92697-2025, USA}
\author{Michael Thorwart}
\affiliation{I. Institut für Theoretische Physik, Universität Hamburg, Notkestraße 9, 22607 Hamburg, Germany}
\affiliation{The Hamburg Centre for Ultrafast Imaging, Luruper Chaussee 149, 22761 Hamburg, Germany}

\date{\today}

\begin{abstract}
We develop a microscopic theory for the two-dimensional spectroscopy of one-dimensional topological superconductors. We consider a ring geometry as a realization of the Kitaev chain with periodic boundary conditions. We show numerically and analytically that the cross-peak structure of the 2D spectra carries unique signatures of the topological phases of the chain. Our work reveals how 2D spectroscopy can identify topological phases in bulk properties, bypassing energy-specific differences caused by topologically protected or trivial boundary modes that are otherwise hard to distinguish.
\end{abstract}

\maketitle

Topological phases of matter have attracted considerable attention following the discovery of topologically non-trivial magnetic and electronic phenomena like the Berezinskii-Kosterlitz-Thouless transition \cite{Berezinskii1970,Berezinskii1972,Kosterlitz1972,Kosterlitz1973} and the integer and fractional quantum Hall effect \cite{vonKlitzing1980,TsuiStoermer1982}. Some topological systems, such as superconducting quantum wires \cite{Kitaev2001}, spin liquids \cite{Kitaev2006} and vortices on surfaces of topological superconductors \cite{FuKane2008} are predicted to host anyons such as spatially isolated Majorana zero-energy boundary modes that are of interest to quantum information processing \cite{Nayak2008,Alicea2011}. Despite experimental evidence of zero-energy modes \cite{Kouwenhoven2012}, their topological origin remains inconclusive \cite{Kouwenhoven2021}. Experimental techniques that reliably identify one-dimensional topological superconductors are badly needed. Current  approaches detect the localized zero-energy boundary modes, but cannot unambiguously discriminate them against topologically trivial features that appear close to zero energy as well, like Yu-Shiba-Rusinov states \cite{Yu1965,Shiba1968,Rusinov1969,Heinrich2018,Cornils2017}, Kondo peaks \cite{Mueller1971,Moca2021}, Andreev bound states \cite{Yu2021}, and Caroli-de-Gennes-Matricon states \cite{Caroli1964,Wang2018}. In two-dimensional electronic systems, dispersive Majorana edge modes have been shown to increase the linear optical conductivity \cite{Nagaosa}. 

A versatile advanced tool is nonlinear two-dimensional (2D) spectroscopy  \cite{Mukamel1995,Valkunas2013} applied in the THz frequency regime to probe electronic excitations in solid-state nanostructures \cite{KuehnJPCB11,WoernerNJP13,NardinSST16,MarkmannNano20} or the Fermi glass phase in disordered correlated materials \cite{MahmoodNatPhys21}. Recently, 2D spectroscopy of two- and three-dimensional topological spin liquids has theoretically revealed characteristic spectral properties of itinerant spin-based anyons and fractons \cite{WanPRL19,Choi2020,Nandkishore2021} and  of strongly correlated two-band Fermi-Hubbard models \cite{PhucPRB21}. It offers additional features in comparison to pump-probe THz spectroscopy \cite{Elsaesser,Nelson,Heinz,Zanni}. The main difference  lies in the decoupling of the waiting time and excitation frequency resolution both of which are high \cite{Gelzinis19}. This is in stark contrast to pump-probe spectroscopy where both are inherently connected by a Fourier uncertainty. Moreover, the lack of large background signals permits excellent signal-to-noise ratios.

In this Manuscript, we employ 2D nonlinear spectroscopy to analyze the periodic Kitaev chain, the archetype of one-dimensional topological superconductors, describing the topological electronic properties of nanowires \cite{Kouwenhoven2012}, atomic magnetic chains \cite{NadjPerge2014,Kim2018}, and cold atom systems \cite{Ruhman2015}.
Rather than investigating the Majorana boundary modes of this model, we consider a periodic configuration to study the topological properties of the bulk and characterize its two phases by 2D spectroscopy.
This could be realized by atomic chain quantum corrals.
In particular, we compare Kitaev chains with the same bulk energy spectrum but a different topological phase. We predict experimental signatures due to topological effects, eliminating differences caused solely by the bulk energy spectra or topologically trivial or non-trivial localized zero-energy states. 
We find signatures of superconducting topological band inversion in the 2D spectra, which are  characteristic for the topological phase and which are absent linear absorption spectra.
Our predictions should be verifiable by 2D THz spectroscopy \cite{KuehnJPCB11,WoernerNJP13,NardinSST16,MarkmannNano20,MahmoodNatPhys21}.

\begin{figure*}
\subfloat{\includegraphics[width=.31\textwidth]{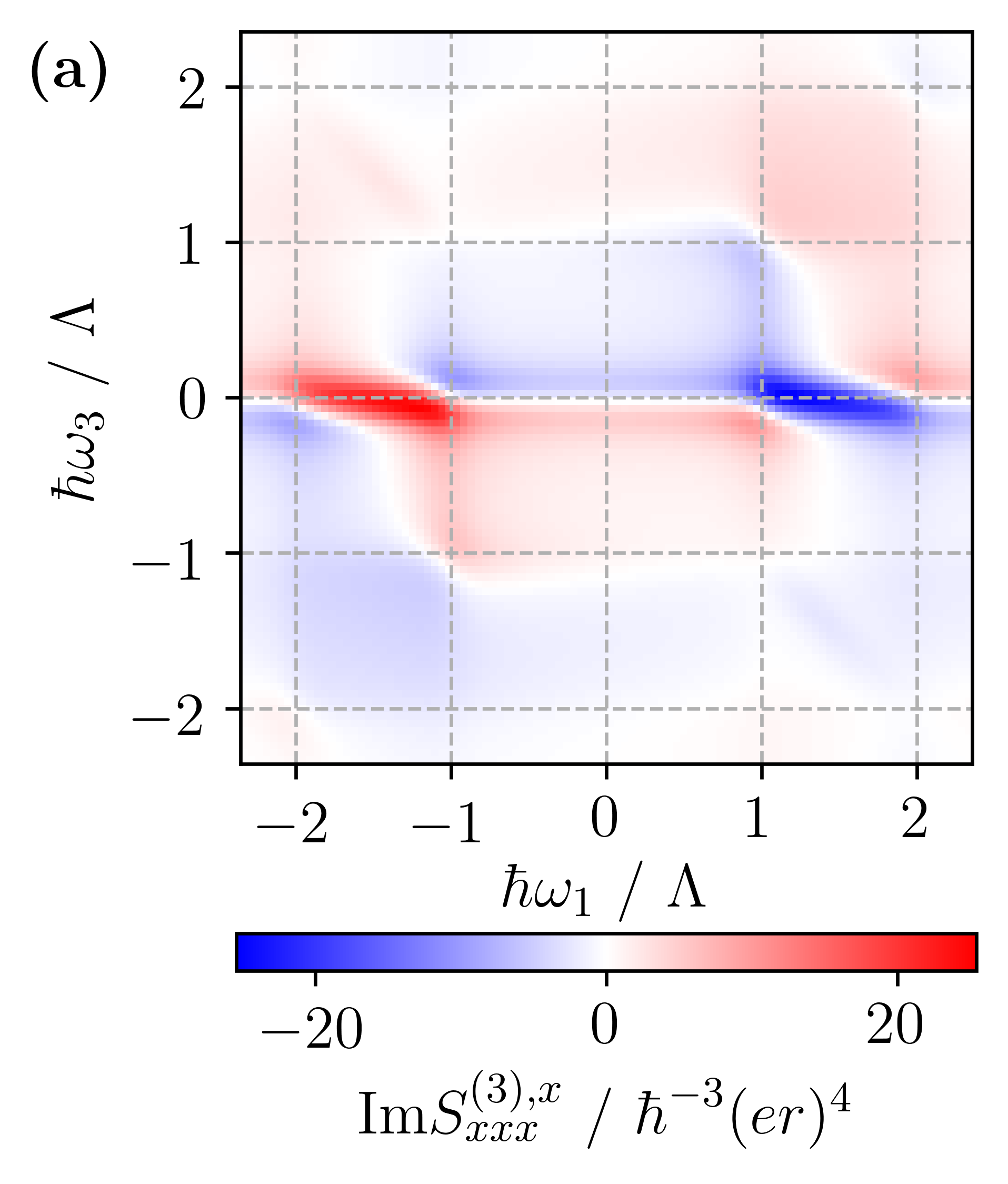}} \hfill
\subfloat{\includegraphics[width=.31\textwidth]{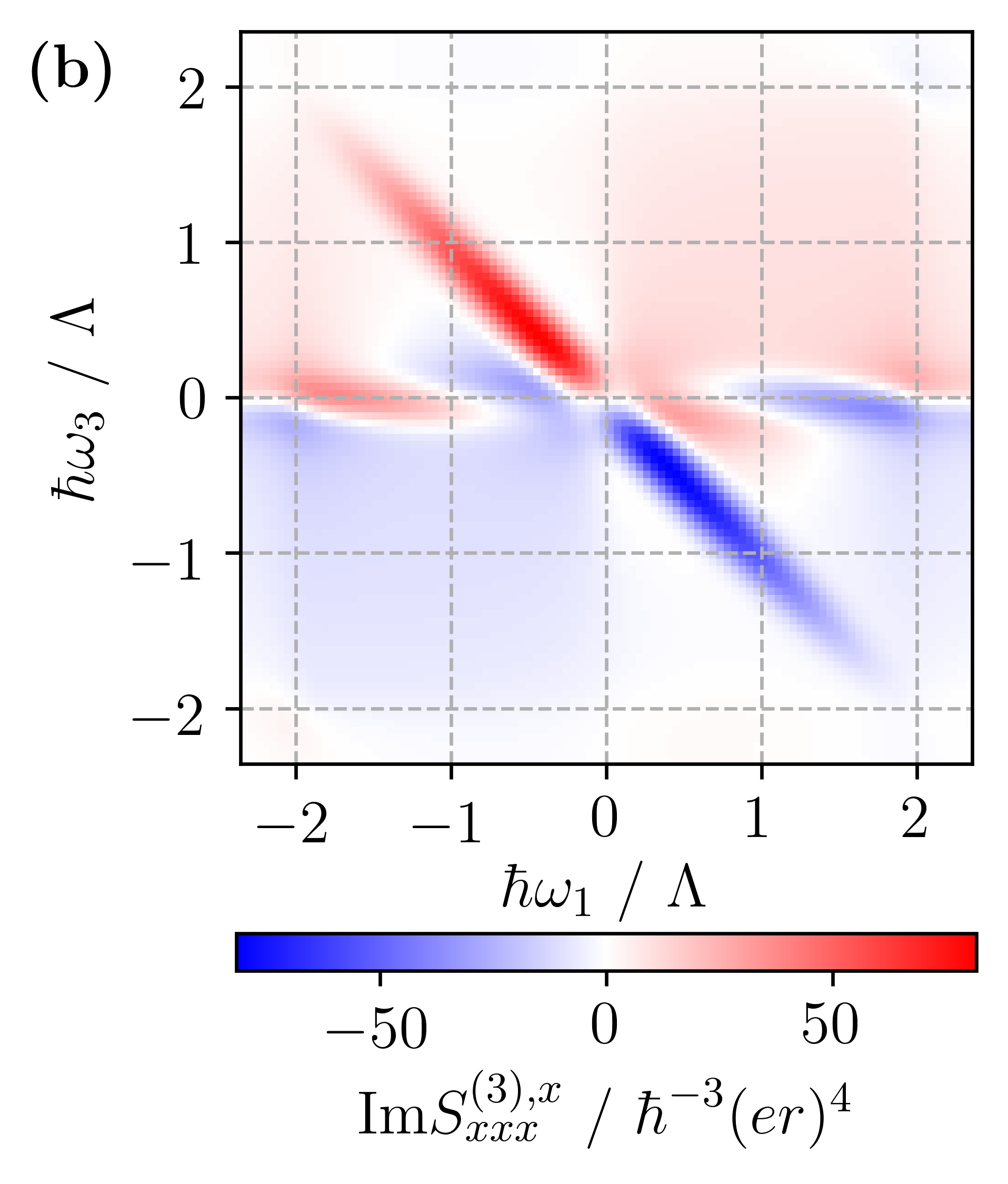}} \hfill
\subfloat{\includegraphics[width=.31\textwidth]{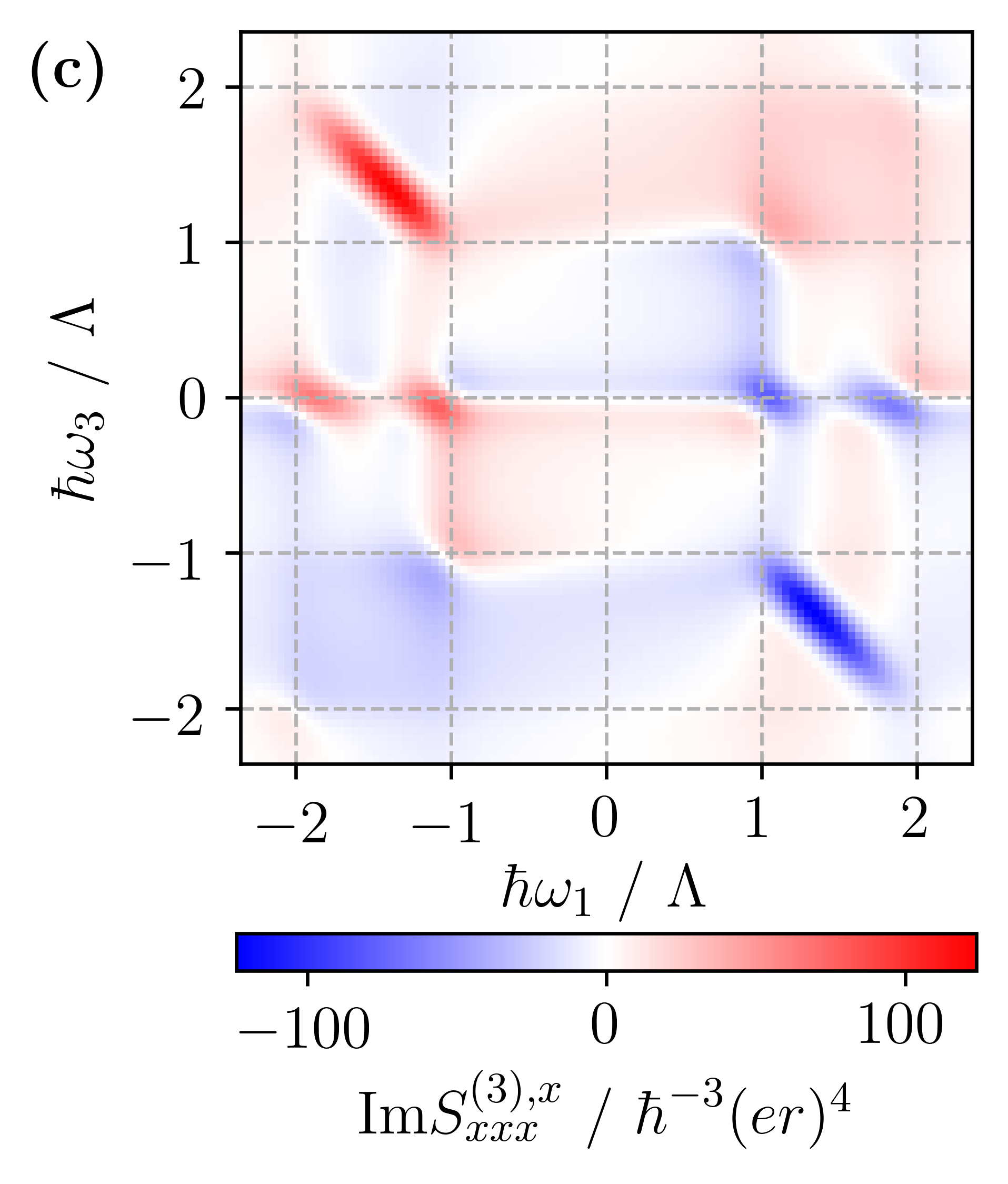}}
\caption{\label{fig:half_gapped}Imaginary part of the 2D spectrum of the Kitaev ring at waiting time $t_2=0$ (a) in the topologically trivial phase with  $\mu= 0.375\Lambda, w = \Delta = 0.125\Lambda$, (b) at the critical point in-between with $\mu= 0.25\Lambda, w = \Delta =  0.25\Lambda$, and (c) in the non-trivial phase with $\mu= 0.125\Lambda, w = \Delta = 0.375\Lambda$.  The chain length is $N=60$, $\Lambda$ is the maximal excitation energy of a single quasiparticle. The topologically trivial and nontrivial phases are distinguished by peaks on the counterdiagonal and the splitting of the peak on the horizontal.}
\end{figure*}

\paragraph*{Model.---}
The Kitaev chain is a one-dimensional  spin-polarized unconventional superconductor with the Hamiltonian
\begin{equation}
\begin{split}
\label{eq:KitaevHamiltonian}
    H = \sum_{n=1}^N \left[ -w a_{n+1}^\dagger a_n - \mu a_n^\dagger a_n + \Delta a_n a_{n+1} \right] + \text{h.c.},
\end{split}
\end{equation}
where $a_n$ is a fermionic annihilation operator, $2\mu$ is the chemical potential, $w$ the nearest-neighbor hopping, and $\Delta$ is the complex superconducting gap parameter   \cite{Kitaev2001}. In physical systems, the parameters can assume a wide range of energies starting from suspended hybridizing atomic chains or semiconductors where they are of the order of eV and going down to meV in hybridized Yu-Shiba-Rusinov states \cite{Kouwenhoven2012,Schneider2021}. However, the superconducting gap is always in the meV range or less.
The system has an electronic band gap for $\abs{w}\neq\abs{\mu}$ and $\Delta \neq 0$ \cite{Kitaev2001}.
For dominant hopping $\abs{w}>\abs{\mu}$, the open chain, i.e., $a_{N+1} = 0$, has an in-gap mode localized at both ends of the chain. Its energy is exponentially small in the system size. In the large-$N$ limit, this mode decomposes into two spatially isolated Majorana operators \cite{Kitaev2001} whose existence is protected by the electronic energy gap in the bulk. The mode can only disappear by closing the gap. Hence, there are two distinct gapped phases: the topologically trivial phase without  and the topologically non-trivial phase with Majorana end modes. Both are characterized by a $\mathbb{Z}_2$ topological invariant of the bulk only \cite{Kitaev2009,Note0}. The boundary modes are due to an interface between different topological phases explained by the bulk boundary correspondence \cite{Essin2011}.

Kitaev \cite{Kitaev2001} has already pointed out that there is a map in form of a simple parameter transformation that leaves the band structure of the periodic chain invariant but changes the topological phase.
We find that the transformed parameters are given by 
\begin{equation}
\label{eq:invariantmap}
    \mu^\prime = \pm w, \ w^\prime = \pm \mu, \ \Delta^\prime = e^{i \vartheta} \sqrt{\mu^2 + \abs{\Delta}^2 - w^2},
\end{equation}
where  $\vartheta$ is an arbitrary real number.
If the system is originally in the non-trivial phase, i.e., $\abs{\mu} < \abs{w}$, then the transformed chain with the primed parameters will be in the trivial phase, because $\abs{w^\prime} = \abs{\mu} < \abs{w} = \abs{\mu^\prime}$.
The same holds vice versa. By this, a dual Hamiltonian with the same spectrum but the inverse topological phase is assigned to each topologically trivial one. Yet, if $\mu^2 + \abs{\Delta}^2 - w^2 < 0$, which can only happen in the non-trivial phase, there will be no trivial Hamiltonian with the same band structure.

We start with the simplest case, $w = \Delta$.
The linear transformation $U$ defined by
\begin{equation}
\label{eq:bogtrafo}
	U^\dagger a_n U = i \left( a_n^\dagger - a_n - a_{n+1}^\dagger - a_{n+1}\right) / 2
\end{equation}
corresponds to the transformed parameters $\mu^\prime=w$ and $w^\prime= \Delta^\prime =\mu$.
In general, we can construct the map between the phases by concatenating the Bogoliubov transformation diagonalizing the trivial Hamiltonian with the inverse of the transformation that diagonalizes the non-trivial Hamiltonian with the same band structure.
Even simpler, the map in Eq.~(\ref{eq:bogtrafo}) can be extended to $\abs{w} \leq \abs{\Delta}$ by fixing the superconducting phase to $\varphi = \arccos(w / \abs{\Delta})$.

\begin{figure*}
\subfloat{\includegraphics[width=.4\textwidth]{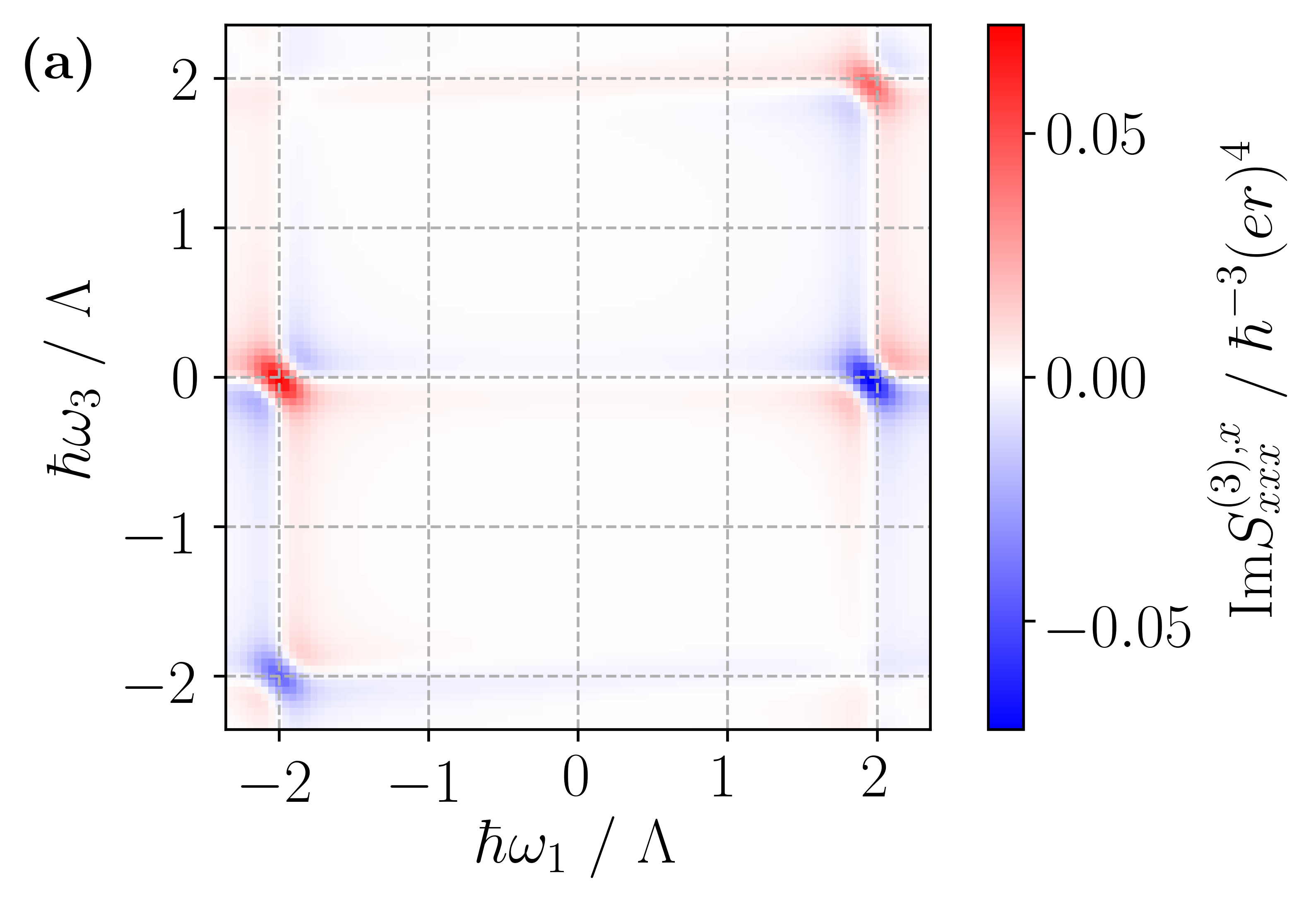}} \hspace{1cm}
\subfloat{\includegraphics[width=.4\textwidth]{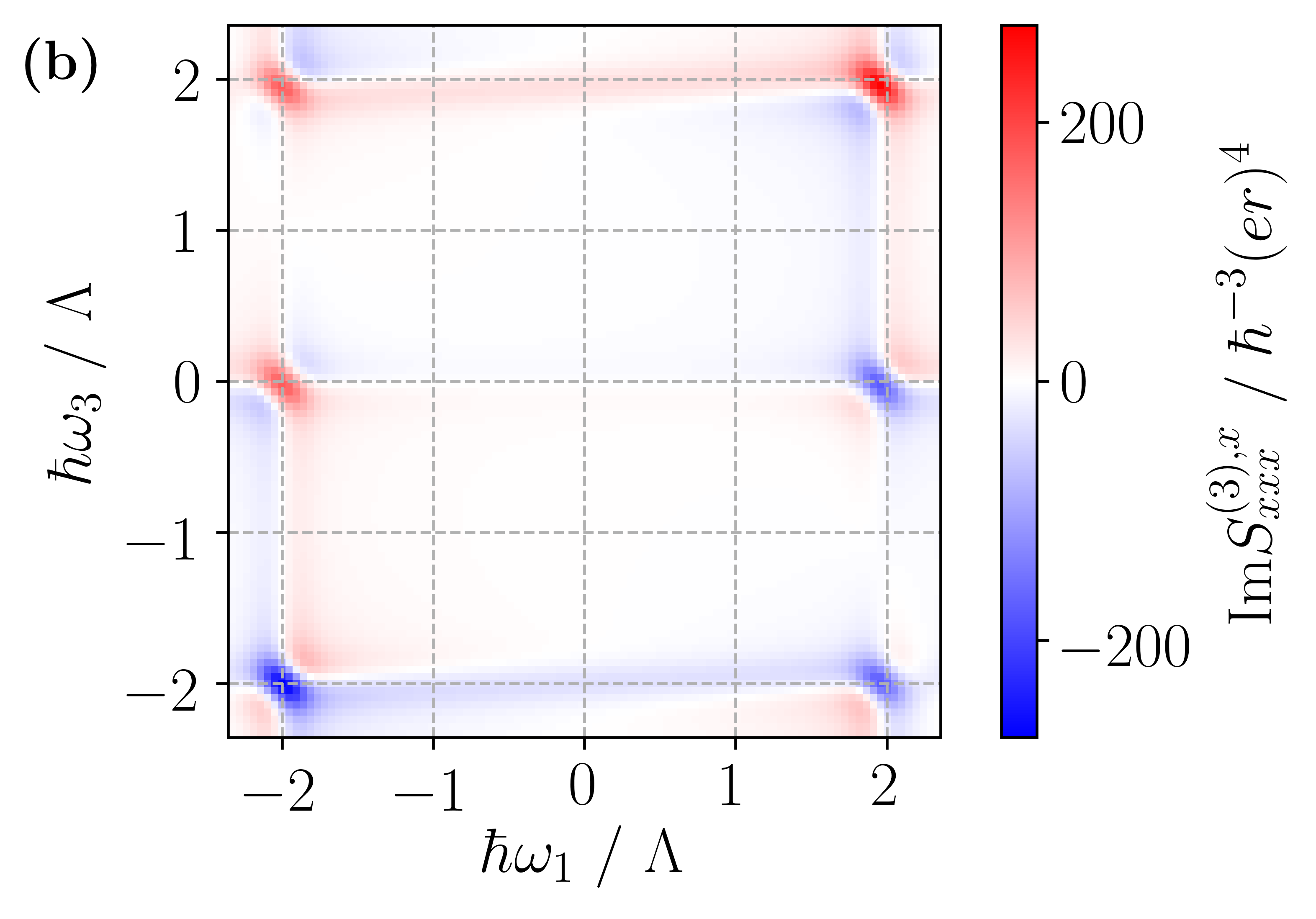}}
\caption{\label{fig:flat_bands}
Imaginary part of the 2D spectrum of the Kitaev ring in (a) the topologically trivial phase with $\mu= 0.005\Lambda, w = \Delta = 0.495\Lambda$, and (b) the non-trivial phase with  $\mu= 0.495\Lambda$ and $w = \Delta = 0.005\Lambda$ for $N=60$.}
\end{figure*}

\paragraph*{2D spectroscopy.---}

In 2D spectroscopy, the system is subjected to three consecutive electromagnetic pulses and its response is probed by interference with a fourth pulse \cite{Mukamel1995,Valkunas2013}. In the dipole approximation, i.e., when  the shortest wavelength of the light is much larger than the extent of the chain, the radiation-matter interaction Hamiltonian reads
$ V(t) = - \mathbf{d} \cdot \mathbf{E}(t)$,
where $\mathbf{d}$ denotes the dipole operator and $\mathbf{E}(t)$ the electric field.
For the Kitaev chain, $\mathbf{d} = -e \mathbf{R}$ with the position operator $\mathbf{R} = \sum_{n=1}^N \mathbf{r}_{n} a^\dagger_n a_n$ and  $e$ the electron charge.
Here, $\mathbf{r}_{n}$ is the location of site $n$. We consider a ring of radius $r$ with 
$\mathbf{r}_{n} =  r \begin{pmatrix} \cos\left(2\pi n/N\right),  \sin\left(2\pi n/N\right), 0 \end{pmatrix}^T$. A similar dipole operator emerges from a low-energy description of realistic systems as shown for a Rashba wire in Sec.~\ref{sec:Rashba} of the Appendix.

We are interested in the time-dependent polarization  $\mathbf{P}(t) = \langle \mathbf{d}(t) \rangle_{\rho(t)}$ which provides the measurable electromagnetic response. Here, $\rho(t)$ is the density matrix of matter. Because the system consists of broad electronic bands, we compute the full third-order signal $\mathbf{P}^{(3)}(t)$ for the 2D spectra which is the sum of all phase matching directions. It can be detected in a collinear beam geometry. Breaking it into phase matching components could reveal additional information on specific groups of dynamical pathways, which goes beyond the present study. Coherent 2D techniques, in particular the double quantum coherence, are usually applied to discrete electronic systems like molecules \cite{Abramavicius2009}. 

We assume that at time $t=0$ the system is in its ground state, and obtain the third-order contribution to the polarization \cite{Mukamel1995,Valkunas2013}
\begin{align}
\begin{split}
    P^{(3), j}(t) =& \int\limits^\infty_{0} dt_\text{3} dt_\text{2} dt_{1} E^{m}(t-t_3) E^{l}(t-t_3-t_{2}) \\
    &\times E^{k}(t-t_3-t_{2}-t_{1}) S^{(3), j}_{klm}(t_3, t_2,t_1) ,
\end{split}
\end{align}
with a sum over repeated indices and the third-order response function $S^{(3), j}_{klm}\left(t_{3}, t_{2}, t_{1}\right)$. The 2D signal is displayed by its Fourier transform with respect to $t_1$ and $t_3$ as 
\begin{align}
\begin{split}
\label{eq:thirdorder}
&S^{(3), j}_{klm}\left(\omega_{3}, t_{2}, \omega_{1}\right) = \frac{2}{\hbar^3} \theta\left(t_{2}\right) \\
&\times
\sum_{\alpha=1}^{4} \int\limits^\infty_{0} \int\limits^\infty_{0} \operatorname{Im} C_{\alpha , klm}^j\left(t_{3}, t_{2}, t_{1}\right) e^{i(\omega_1 t_1 + \omega_3 t_3)} dt_1 dt_3,
\end{split}
\end{align}
with the Heaviside function $\theta(t)$ and $C_\alpha$ are the four-point correlation functions of the dipole operator (see Sec.~\ref{sec:corrFunctions} of the Appendix).
$\omega_{1/3}$ is the excitation/detection frequency and $t_2$  the waiting time. In the following, we set $t_2=0$.

\paragraph*{Results.---}
We restrict the discussion to the $S^{(3), x}_{xxx}$-component, where all light pulses are polarized in the $x$-direction.
The signals for this feasible configuration are similar to the ones for a physically unrealistic linear chain with periodic boundary conditions.
We choose a representative slice in the $\left(w=\Delta\right)$-plane to demonstrate the parameter dependence of the 2D spectra. By this,  we can use the map in Eq.~(\ref{eq:bogtrafo}) to clarify the qualitative differences between the phases.
Representatives of the two phases are the trivial atomistic limit (dissected atoms) and the sweet spot of the Majorana chain that hosts localized Majorana modes in an open chain. We fix the maximal quasiparticle energy $\Lambda$ as the energy scale. In our case, $\Lambda$ can be in the meV regime, but depending on the physical system, $\Lambda$ can vary up to eV \cite{Kouwenhoven2012,Schneider2021}. We follow the trajectory 
\begin{align}
\label{eq:paramcurve}
    \Gamma_s = \left( \mu_s, w_s, \Delta_s \right)
    = \Lambda \left( 1-s, \ s, \ s \right) / 2,{}
\end{align}
where $0 \leq s \leq 1$, which  interpolates between the two extreme cases $H(\Gamma_0)$ being the Hamiltonian in the atomistic limit and $H(\Gamma_1)$  the Hamiltonian for the sweet spot, such that $\Lambda$ remains unchanged at all instances. For $s < 0.5$, $H(\Gamma_s)$ is in the topologically trivial phase, for $s > 0.5$ in the non-trivial phase, and for $s=0.5$, the system reaches the semi-metallic critical point, where the bulk gap closes. The spectra and band structures of $H(\Gamma_s)$ and $H(\Gamma_{1-s})$ coincide due to the map $U$ in Eq.~(\ref{eq:bogtrafo}). By this, 2D spectra for different topological phases with the same eigenenergies can be compared.

\begin{figure*}
\subfloat{\includegraphics[scale=0.5]{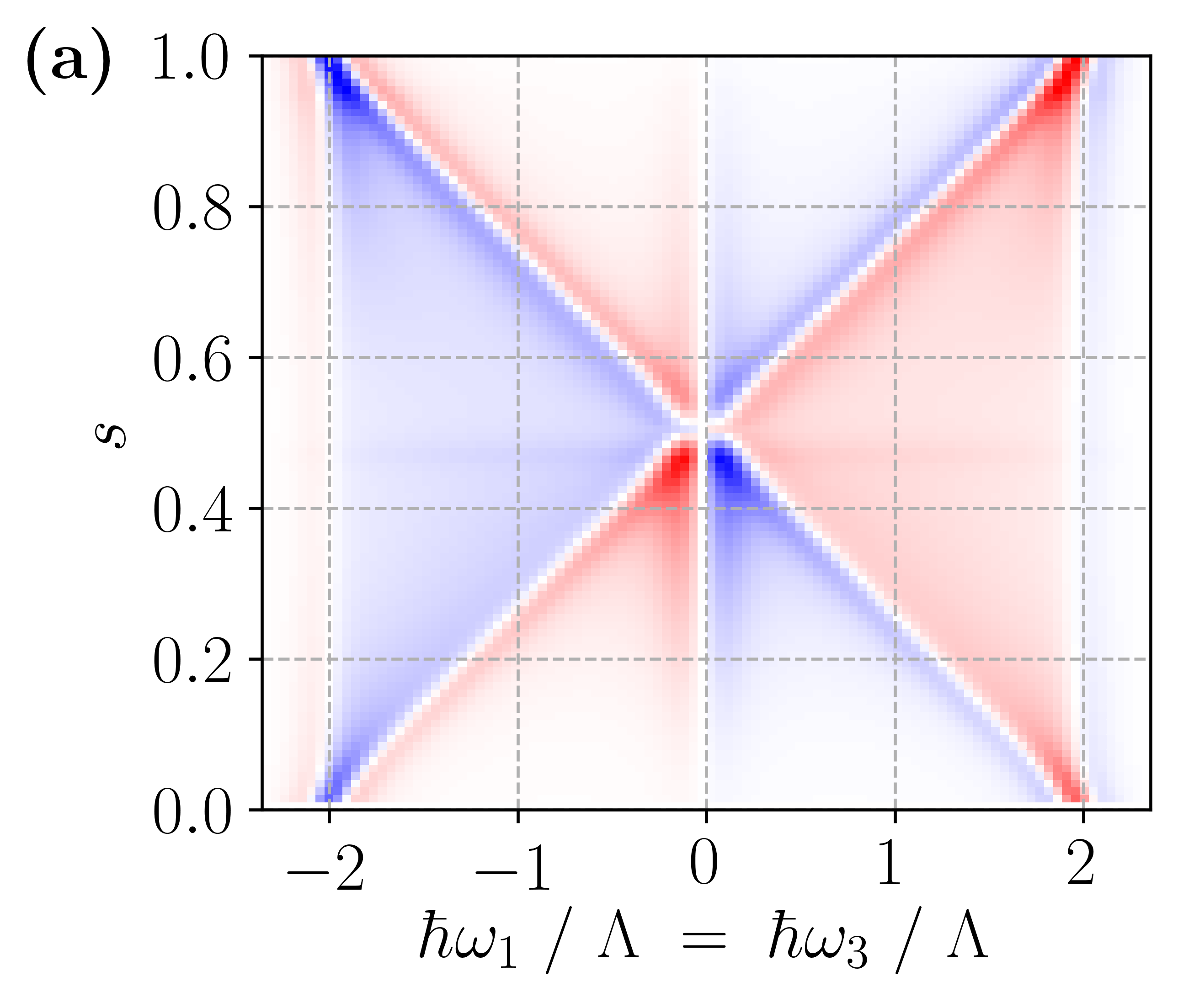}}
\subfloat{\includegraphics[scale=0.5]{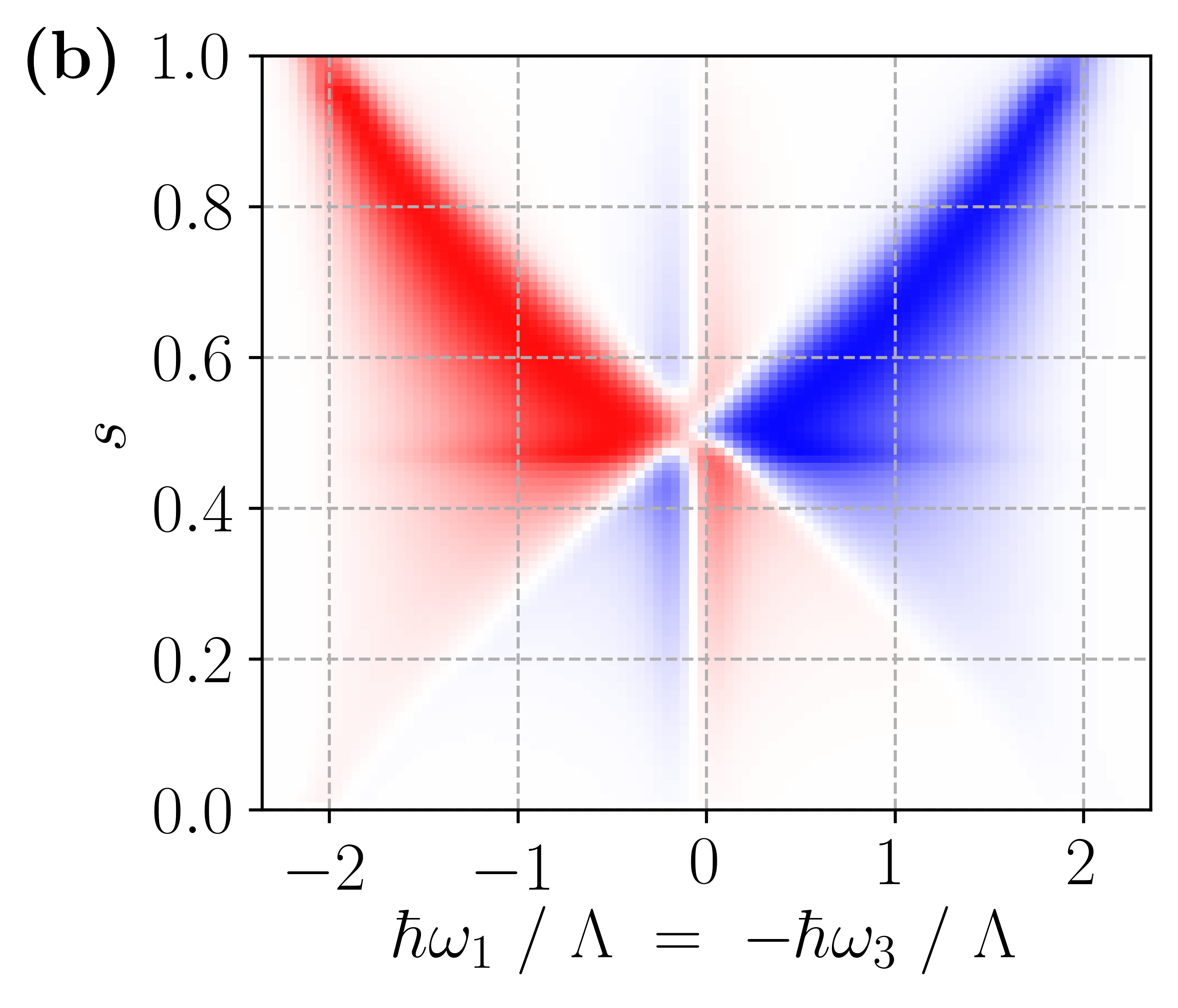}}
\subfloat{\includegraphics[scale=0.5]{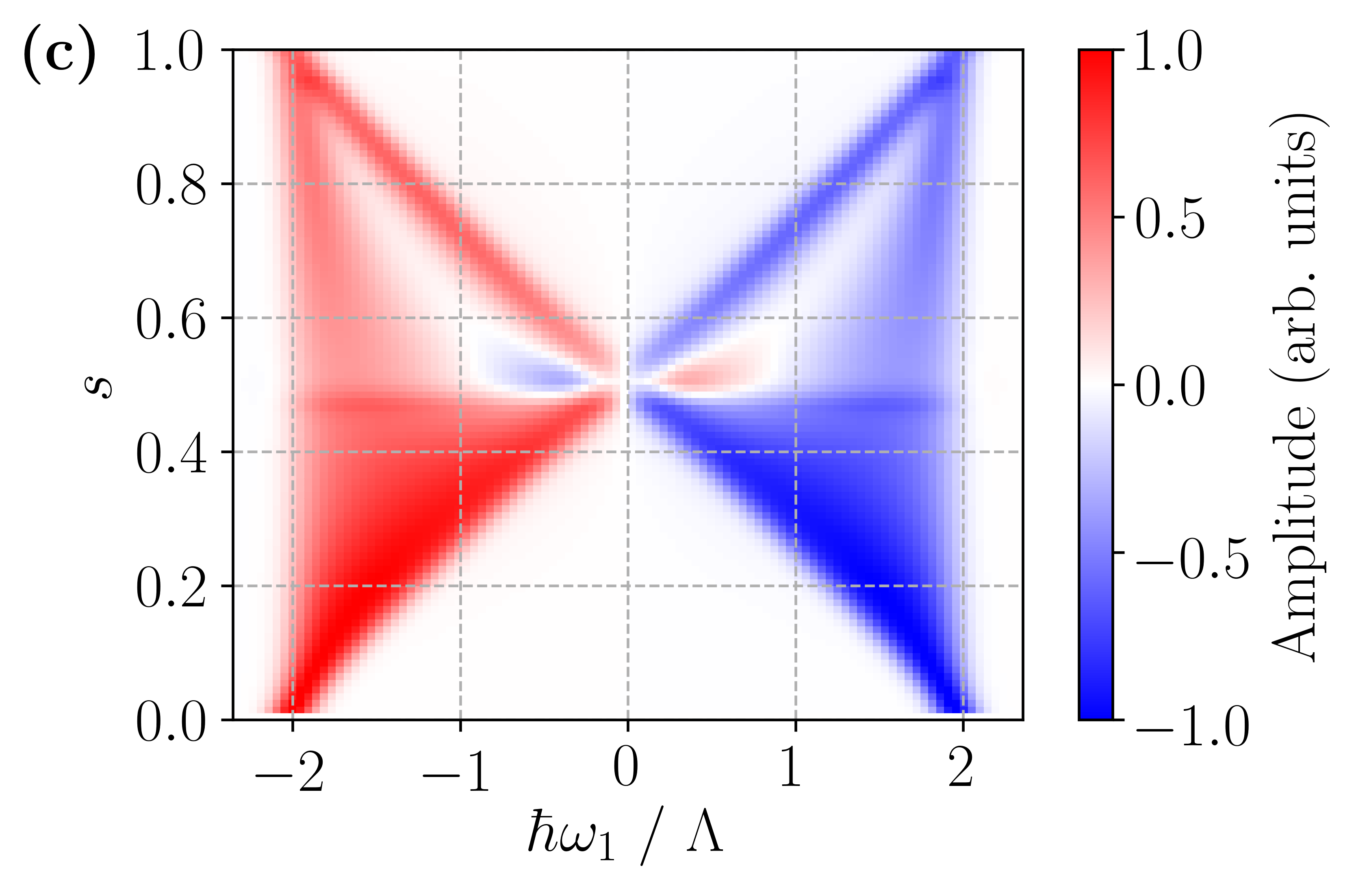}}
\caption{\label{fig:sections_ring}(a) Diagonal, (b) counterdiagonal and (c) horizontal sections of the imaginary part of the 2D spectra for the Hamiltonian $H(\Gamma_{s})$ as a function of $s$ following Eq.~(\ref{eq:paramcurve}). For each parameter set $\Gamma_s$, the 2D spectra are normalized to their maximal peak amplitude. The chain length is $N=60$. Differences between the topological phases emerge along the counterdiagonal and the horizontal lines. In the trivial phase $\left(s<0.5\right)$, the counterdiagonal peaks disappear. The horizontal peaks are more pronounced in the trivial phase than in the non-trivial phase $\left(s>0.5\right)$.}
\end{figure*}

Representative 2D spectra for a band gap of $\Lambda / 2$ and for the gapless critical point are shown in Fig.~\ref{fig:half_gapped} (see also the Supplemental Movie \cite{Note2}). They include a Gaussian broadening ($\sigma=0.05\Lambda$) to increase readability.
Noticeable peaks in the 2D spectra are arranged along three main axes, the diagonal $\omega_1 = \omega_3$, the counterdiagonal $\omega_1 = -\omega_3$, and the horizontal $\omega_3 = 0$. Valuable information is contained in the cross-peaks on the counterdiagonal and the horizontal.
A change of the cross-peak pattern is observed when passing from the topologically trivial to the non-trivial phase. The counterdiagonal peaks dominate the non-trivial phase, while they almost disappear in the trivial phase. The horizontal peaks appear in both phases. They form a large inhomogeneously broadened peak in the trivial phase but become disconnected in the non-trivial phase and are most pronounced at the band edges. Furthermore, their relative magnitude significantly decreases.
In general, the overall magnitude of the 2D spectra increases for $s\rightarrow1$. The peak amplitudes between the phases differ by orders of magnitude. For perfectly flat bands in the trivial phase, it can even vanish due to the charge conserving nature of the dipole operator. The ground state in the trivial phase with flat bands is either the empty or fully filled lattice. There are no other states with the same charge, hence, all transitions are forbidden. For flat bands in the non-trivial phase, there are numerous possible transitions, in contrast. The charge expectation value of the ground state is $-Ne/2$. We estimate that for even $N$, the number of Fock states with charge $-Ne/2$ is $2^N / \sqrt{N}$ due to Sterling's formula. This accounts for the discrepancy of the magnitudes between the 2D spectra of the almost flat band scenarios shown in Fig.~\ref{fig:flat_bands}.

For the nearly flat bands in Fig.~\ref{fig:flat_bands}, we find essential differences between the 2D spectra of the two topological phases. In the trivial phase, the horizontal peaks are the dominant cross-peaks while counterdiagonal peaks are absent. In the non-trivial phase, the counterdiagonal peaks are dominant while the horizontal peaks are reduced. To show that this is generic, we depict the cross sections along the diagonal, counterdiagonal and horizontal in Fig.~\ref{fig:sections_ring}. For each value of $s$, the 2D spectra are normalized to their maximal peak amplitude. The diagonal at $t_2 = 0$ carries information on the linear response spectra. 
We find the 2D spectra to be symmetric about the phase transition at $s=0.5$. This reaffirms that the linear response cannot uncover differences between the phases. 
Our analytic calculations show that the difference between the phases in linear spectroscopy is essentially a scaling factor \cite{Note1}.
For the counterdiagonal, cross-peaks disappear in the trivial regime $s < 0.5$, but are strong in the non-trivial regime $s > 0.5$. 
Importantly, the change in the relative peak amplitudes when crossing the critical point $s=0.5$ is continuous. 
The signal from the horizontal sections
forms a broad continuum in the trivial phase that is clearly split in the topological phase. This is caused by the superconducting topological band inversion characteristic for the model. The anomalous term in Eq.~(\ref{eq:KitaevHamiltonian}) mixes the particle and hole bands. In the trivial phase, the bands maintain their predominant particle and hole character, respectively. In the topological phase, the bands change between  particle and hole character at the inversion points in the Brillouin zone. There, the non-vanishing two-particle to two-particle transition dipole moments have a gap closure \cite{Note1}.
This is absent in the trivial phase and is thus unique to the topological phase. For large $N$, their transition frequencies go to zero. Hence, they contribute to the horizontal peaks in the 2D spectra, and the observed splitting of the peak continua provides a clear signature of the superconducting topological band inversion.
The difference in the cross-peaks and the absence of any difference in the diagonal peaks are a fundamental advantage of nonlinear spectroscopy for characterizing topological phases.
Our results transfer to finite Kitaev chains with open boundary conditions. Yet, additional Majorana end modes as well as possible trivial zero-energy modes result in a doubling of the 2D spectrum at energies of the order of the band gap that must be accounted for.
Remarkably, the bulk contribution is qualitatively the same as for the periodic configuration, suggesting that our results are largely insensitive to the specific geometry underlying the dipole operator.

The map $U$ offers an alternative interpretation of our results. 
Rather than considering $U$ to actively change the topological phase, we could equivalently consider the Hamiltonian to be invariant and passively transform the measurement operator, i.e., the dipole operator, which has the form of a local chemical potential, into the Majorana braiding operator $B_{n,n+1} = a_{n+1}^\dagger a_n + a_{n+1} a_n + \text{h.c.}$ for adjacent sites \cite{Ivanov2001}. Formally, this means $U^\dagger \mathbf{d} U = \frac{e}{2} \sum_{n=1}^N \mathbf{r}_{n} B_{n,n+1}$.
Then, the 2D spectrum can be interpreted in two ways: first, the chain being in one phase and probed by the common dipole operator, and second the chain being in the other phase and probed by the braiding operator.

\paragraph*{Conclusions.---}

With the Kitaev ring, we propose a physical realization of the Kitaev chain with periodic boundary conditions and calculate the THz  response in 2D nonlinear spectroscopy with three parallel polarized field pulses. By a mapping between the topologically trivial and non-trivial phases that changes the phase but not the band structure of the Kitaev Hamiltonian, we identify signatures stemming solely from topological effects and not from the energy spectra.
A superconducting topological band inversion is then detected by cross-peaks in the 2D spectra which underlines the advantage of nonlinear spectroscopy over linear spectroscopy for topological systems.
A band inversion has recently been resolved in scanning tunneling microscope experiments \cite{Schneider2021}, which couples to the local charge rather than the dipole operator. 2D  spectroscopy is less invasive, offers higher spectral resolution and is less prone to dissipation, where any backaction of a macroscopic tip on the quantum system can be excluded. 
A seeming caveat of our approach is that the superconducting gap $\Delta$ should be rather large for the $U$-map to exist. However, our analytic computation of the dipole moments \cite{Note1} suggests that our results carry over to small $\Delta$.
In contrast to topological spin liquids [24, 25], the electronic system at hand can be probed both in its topologically trivial and nontrivial phase, and its topological features are revealed by bulk properties only, omitting the spectroscopy of  hard-to-control  low-energy topological quasiparticles, that interfere with the topological response of the bulk. 
Future research on multiple topological band inversions and multiband models could help to establish a general connection between our findings and the bulk topological invariant.

\begin{acknowledgments}
We thank Hong-Guang Duan for helpful discussions.
M.T. and F.G.\ acknowledge support by the Cluster of Excellence CUI: Advanced Imaging of Matter of the Deutsche Forschungsgemeinschaft (DFG) – EXC 2056 – project ID 390715994.
T.P.\ acknowledges funding by the Deutsche Forschungsgemeinschaft (DFG) project no. 420120155.
S.M.\ gratefully acknowledges the support of the National Science Foundation Grant CHE-1953045.
\end{acknowledgments}

\widetext

\begin{center}
\textbf{\large Appendix \\[3mm] Unique Signatures of Topological Phases in Two-Dimensional THz Spectroscopy}
\end{center}

\setcounter{equation}{0}
\setcounter{figure}{0}
\setcounter{table}{0}
\makeatletter
\renewcommand{\theequation}{S\arabic{equation}}
\renewcommand{\thefigure}{S\arabic{figure}}
\renewcommand{\bibnumfmt}[1]{[S#1]}

\setlength{\marginparwidth}{2cm}

In this Appendix, we give the explicit expressions for the 4-point correlation functions contributing to the third-order response function of the Kitaev chain. We provide details of the numerical evaluation of the correlation functions. The many-particle transition dipole moments are computed numerically and complemented by analytic results for the large-$N$ limit.

\section{\label{sec:corrFunctions}Correlation Functions}

\noindent
The third-order response function is given by 
\begin{equation}
\label{eq:thirdorderS}
S^{(3), j}_{klm}\left(t_{3}, t_{2}, t_{1}\right) = \frac{2}{\hbar^3} \theta\left(t_{1}\right) \theta\left(t_{2}\right) \theta\left(t_{3}\right)
 \sum_{\alpha=1}^{4}\operatorname{Im} C_{\alpha , klm}^j\left(t_{3}, t_{2}, t_{1}\right)
\end{equation}
with the Heaviside step function $\theta(t)$. The four-point correlation functions $C_\alpha$ are given by 
\begin{align}
\label{eq:C1}
    C^j_{1, klm}\left(t_{3}, t_{2}, t_{1}\right)
    =&\langle d_l\left(t_{1}\right) d_m\left(t_{1}+t_{2}\right) d^j\left(t_{1}+t_{2}+t_{3}\right) d_k(0)\rangle_\rho, 
\\
    C^j_{2, klm}\left(t_{3}, t_{2}, t_{1}\right)
    =&\langle d_k(0) d_m\left(t_{1}+t_{2}\right) d^j\left(t_{1}+t_{2}+t_{3}\right) d_l\left(t_{1}\right)\rangle_\rho, 
\\
    C^j_{3, klm}\left(t_{3}, t_{2}, t_{1}\right)
    =&\langle d_k(0) d_l\left(t_{1}\right) d^j\left(t_{1}+t_{2}+t_{3}\right) d_m\left(t_{1}+t_{2}\right)\rangle_\rho,
 \\
 \label{eq:C4}
    C^j_{4, klm}\left(t_{3}, t_{2}, t_{1}\right)
    =&\langle d^j\left(t_{1}+t_{2}+t_{3}\right) d_m\left(t_{1}+t_{2}\right) d_l\left(t_{1}\right) d_k(0)\rangle_\rho.
\end{align}
Here, $d_j$ is the $j$-th component of the dipole operator $\mathbf{d}$ and $\rho$ is the groundstate of the unperturbed system. For a derivation of these expression, we are referring to Chapter 13 of Ref.~\cite{Valkunas2013}.

\section{Numerical Evaluation of the Correlation Functions}

\noindent
We find the eigenmodes of the Kiteav chain in momentum space by a standard Bogoliubov transformation \cite{Kitaev2001},
where we define all quasiparticle energies to be non-negative.
The quasiparticle vacuum $\ket{\Omega}$ and the groundstate of the system then coincide. With this, the 4-point correlation functions for the $x$-components of the dipole operator are of the form
\begin{equation}
\label{eq:C}
C = \bra{\Omega} d_x (\tau_1)d_x (\tau_2)d_x (\tau_3)d_x (\tau_4)\ket{\Omega}.
\end{equation}
Consider a general quadratic operator for the dipole operator 
\begin{equation}
\label{eq:generalOp}
d_x (\tau) = \sum_{i,j=1}^N \left[A_{ij} (\tau) f_i^\dagger f_j + B_{ij} (\tau) f_i^\dagger f_j^\dagger + C_{ij} (\tau) f_i f_j + D_{ij} (\tau) f_i f_j^\dagger\right].
\end{equation}
The matrices $A$, $B$, $C$ and $D$ are obtained either numerically or analytically from the Bogoliubov transformation that diagonalizes the Hamiltonian. To compute the correlation functions, we insert Eq.~(\ref{eq:generalOp}) into Eq.~(\ref{eq:C}). Evaluating the correlation functions reduces to computing the vacuum expectation values of products of creation and annihilation operators. We achieve this combinatorically involved task in a systematic way by using Wick contractions and Wick's theorem. The results are sums of traces of products of the matrices $A$, $B$, $C$ and $D$ at different times $\tau_1$, $\tau_2$, $\tau_3$ and $\tau_4$ that must be evaluated numerically.

\section{\label{sec:matrixElements}Density of States and the Dipole Operator Matrix Elements}

\noindent
Here, we provide the density of states of two- and four-quasiparticle states as well as the corresponding matrix elements of the dipole operator $\mathbf{d}$. We constrain ourselves to its  $x$-component, because the $z$-component vanishes by definition and the $y$-component carries equivalent information in the rotationally invariant system.
We numerically construct the many-particle Fock states directly from the single-particle states of a Kitaev chain of length $N=60$ by standard combinatorics. In total, there are
$1770$ states with two quasiparticles and 
$487635$ with four quasiparticles.
The corresponding density of states is depicted in Figs.~(\ref{figDOS1}-\ref{figDOS4}) for a representative choice of parameters $s$ (see main text), which cover the topologically trivial phase at $s<0.5$, the critical point at $s=0.5$, and the nontrivial phase at $s>0.5$. 
\newlength{\figWidth}
\setlength{\figWidth}{0.3\linewidth}
\begin{figure}
\centering
\raisebox{-1\height}{
    \parbox{\figWidth}{(a) $s=0.01$\\
    \includegraphics[width = 1\linewidth]{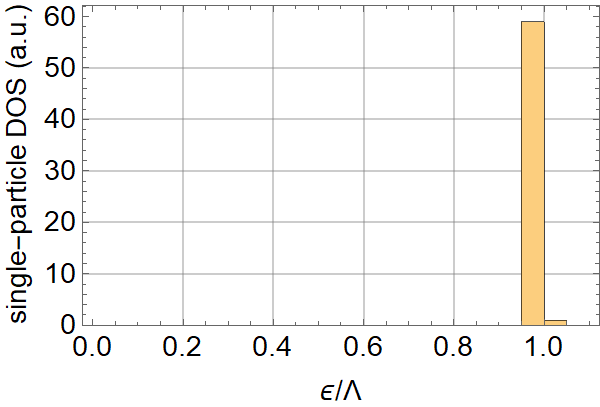}
    }
 }
 \raisebox{-1\height}{
    \parbox{\figWidth}{(b) $s=0.25$\\
    \includegraphics[width = 1\linewidth]{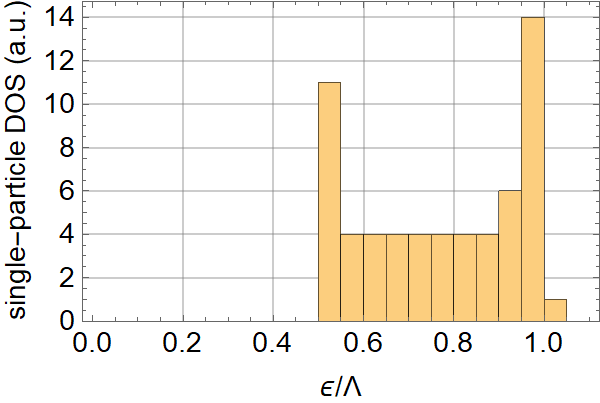}
    }
 }
 \raisebox{-1\height}{
    \parbox{\figWidth}{(c) $s=0.5$\\
    \includegraphics[width = 1\linewidth]{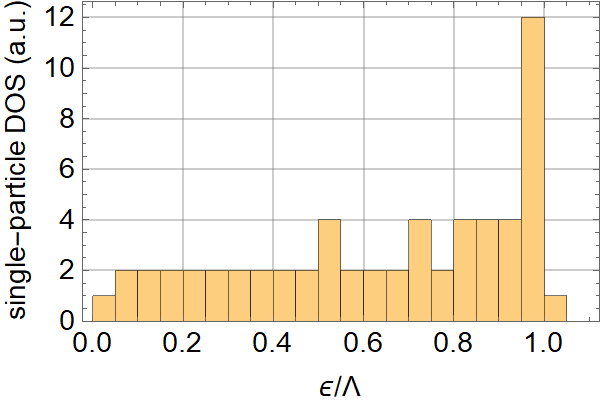}
    }
 }
 \raisebox{-1\height}{
    \parbox{\figWidth}{(d) $s=0.75$\\
    \includegraphics[width = 1\linewidth]{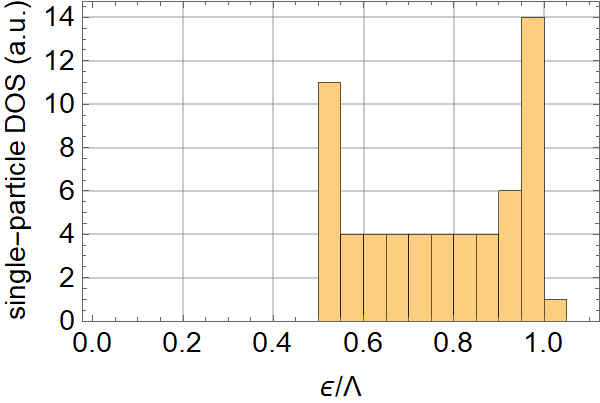}
    }
 }
 \raisebox{-1\height}{
    \parbox{\figWidth}{(e) $s=0.99$\\
    \includegraphics[width = 1\linewidth]{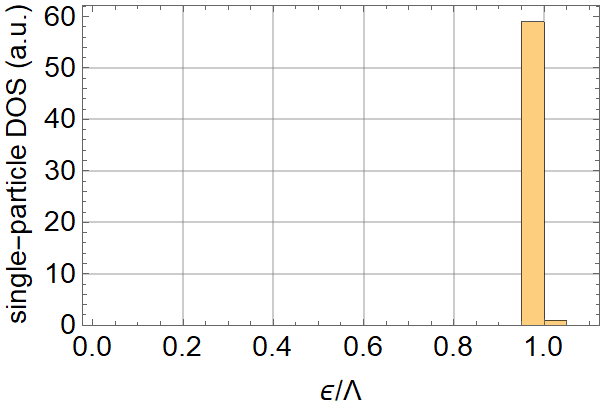}
    }
 }
\caption{\label{figDOS1}Single-particle density of states for the Kitaev chain of length $N=60$ as a general reference at different parameters corresponding to $s=0.01$, $0.25$, $0.5$, $0.75$, $0.99$, respectively (see main text).
} 
\end{figure}
\begin{figure}
 \centering
 \raisebox{-1\height}{
    \parbox{\figWidth}{(a) $s=0.01$\\
    \includegraphics[width = 1\linewidth]{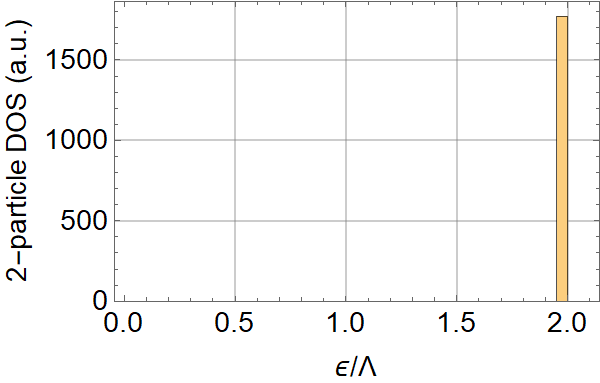}
    }
 }
 \raisebox{-1\height}{
    \parbox{\figWidth}{(b) $s=0.25$\\
    \includegraphics[width = 1\linewidth]{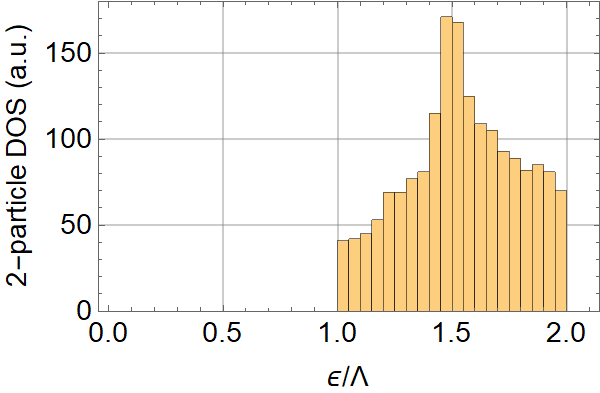}
    }
 }
 \raisebox{-1\height}{
    \parbox{\figWidth}{(c) $s=0.5$\\
    \includegraphics[width = 1\linewidth]{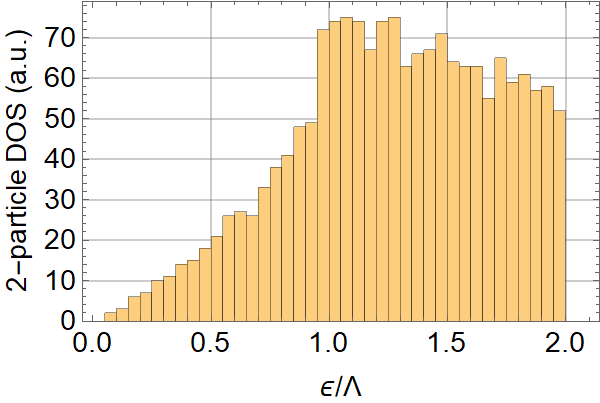}
    }
 }
 \raisebox{-1\height}{
    \parbox{\figWidth}{(d) $s=0.75$\\
    \includegraphics[width = 1\linewidth]{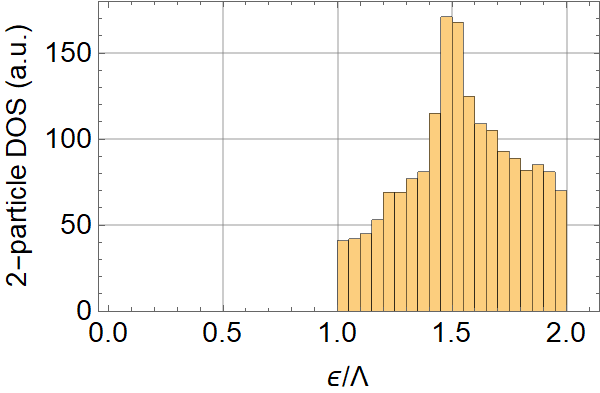}
    }
 }
 \raisebox{-1\height}{
    \parbox{\figWidth}{(e) $s=0.99$\\
    \includegraphics[width = 1\linewidth]{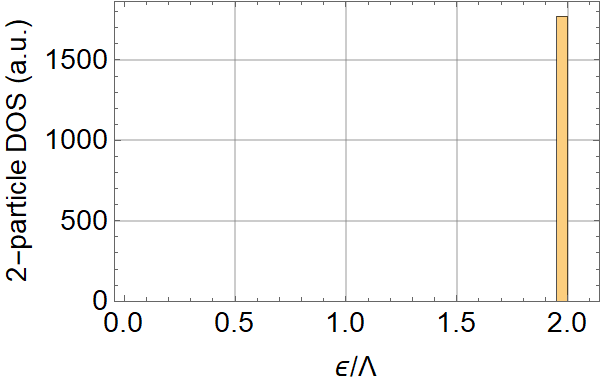}
    }
 }
 \caption{\label{figDOS2}Histograms of the two-particle density of states for chain length $N=60$.}
\end{figure}

\begin{figure}
 \centering
 \raisebox{-1\height}{
    \parbox{\figWidth}{(a) $s=0.01$ \\
    \includegraphics[width = 1\linewidth]{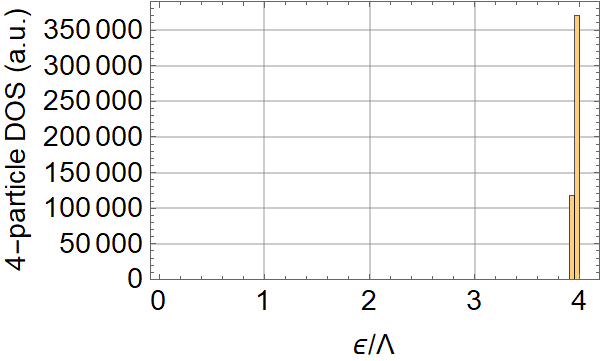}
    }
 }
 \raisebox{-1\height}{
    \parbox{\figWidth}{(b) $s=0.25$\\
    \includegraphics[width = 1\linewidth]{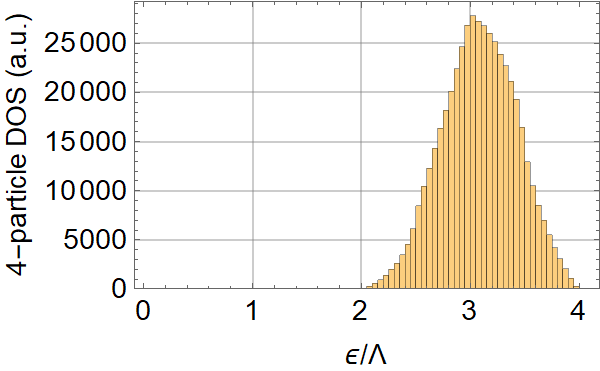}
    }
 }
 \raisebox{-1\height}{
    \parbox{\figWidth}{(c)  $s=0.5$\\
    \includegraphics[width = 1\linewidth]{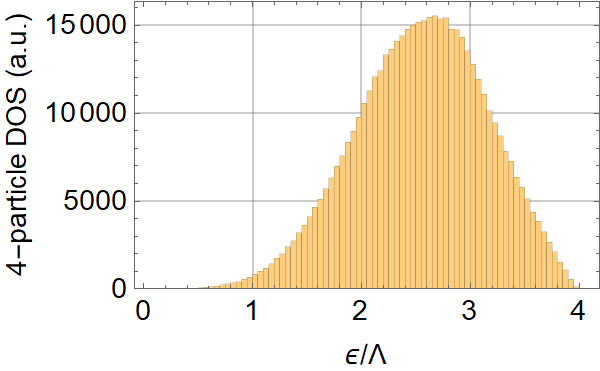}
    }
 }
 \raisebox{-1\height}{
    \parbox{\figWidth}{(d)  $s=0.75$\\
    \includegraphics[width = 1\linewidth]{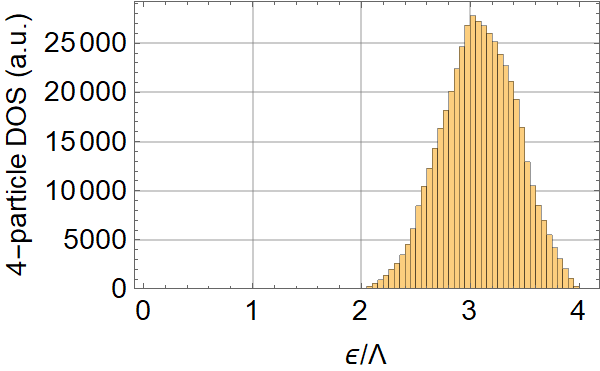}
    }
}
 \raisebox{-1\height}{
    \parbox{\figWidth}{(e)  $s=0.99$\\
    \includegraphics[width = 1\linewidth]{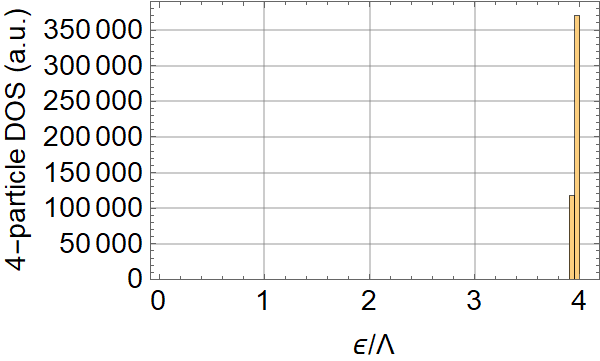}
    }
 }
\caption{\label{figDOS4}Histograms of the four-particle density of states for chain length $N=60$.}
\end{figure}

\begin{figure}
 \centering
 \raisebox{-1\height}{
    \parbox{\figWidth}{(a) $s=0.01$\\
    \includegraphics[width = 1\linewidth]{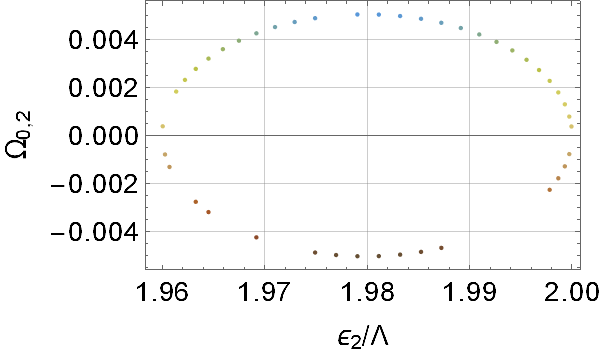}
    }
 }
 \raisebox{-1\height}{
    \parbox{\figWidth}{(b) $s=0.25$\\
    \includegraphics[width = 1\linewidth]{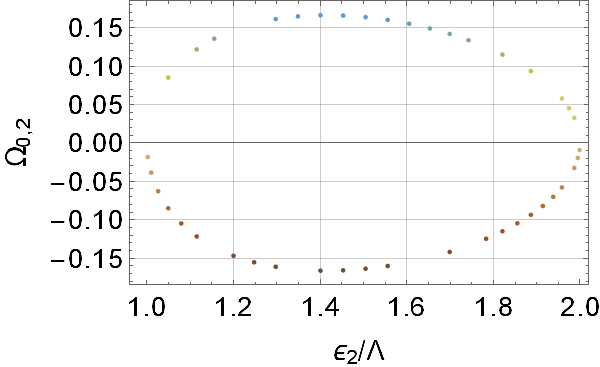}
    }
 }
 \raisebox{-1\height}{
    \parbox{\figWidth}{(c) $s=0.5$\\
    \includegraphics[width = 1\linewidth]{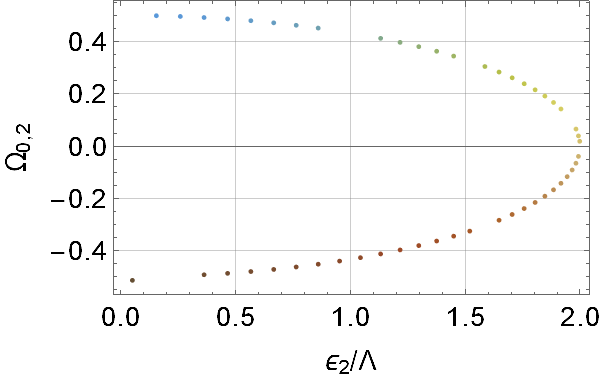}
    }
 }
 \raisebox{-1\height}{
    \parbox{\figWidth}{(d) $s=0.75$\\
    \includegraphics[width = 1\linewidth]{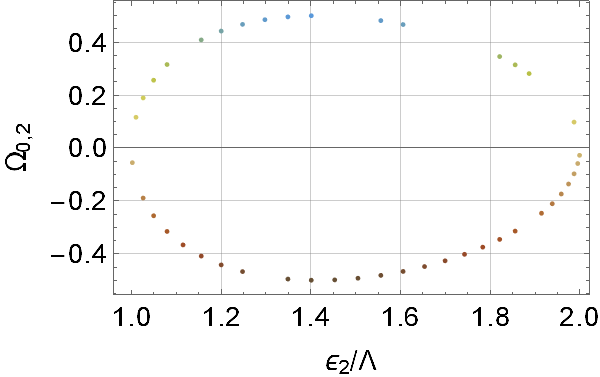}
    }
 }
 \raisebox{-1\height}{
    \parbox{\figWidth}{(e) $s=0.99$\\
    \includegraphics[width = 1\linewidth]{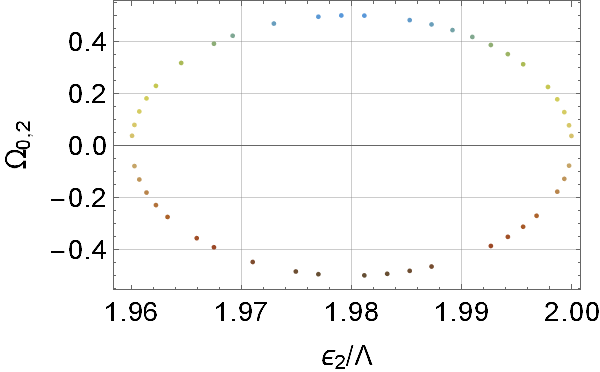}
    }
 }
 
\caption{\label{figDipoleMatrixElements}Dipole matrix elements $\Omega_{0,2}$ between the ground state and states with two quasi-particles excitations and energy $\epsilon_2$.}
\end{figure}

\begin{figure}
 \centering
 \raisebox{-1\height}{
    \parbox{\figWidth}{(a) $s=0.01$\\
    \includegraphics[width = 1\linewidth]{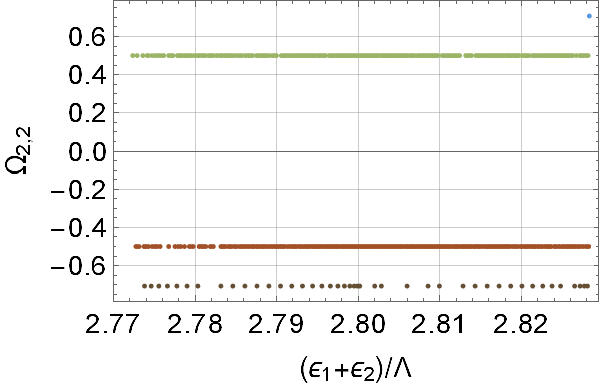}
    }
 }
 \raisebox{-1\height}{
    \parbox{\figWidth}{(b) $s=0.25$\\
    \includegraphics[width = 1\linewidth]{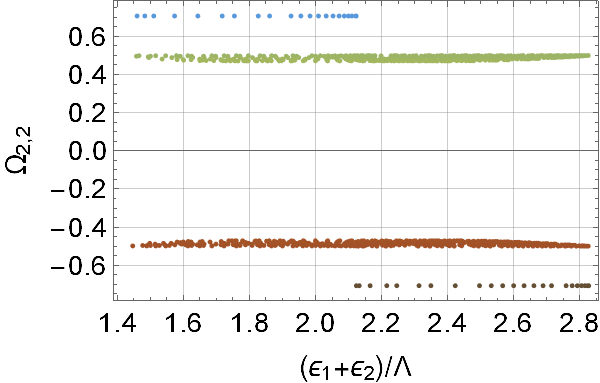}
    }
 }
 \raisebox{-1\height}{
    \parbox{\figWidth}{(c) $s=0.5$\\
    \includegraphics[width = 1\linewidth]{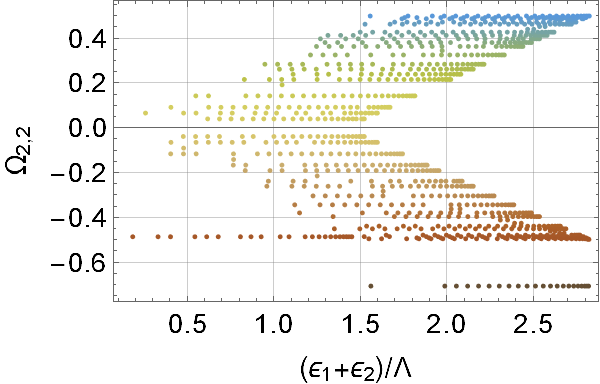}
    }
 }
 \raisebox{-1\height}{
    \parbox{\figWidth}{(d) $s=0.75$\\
    \includegraphics[width = 1\linewidth]{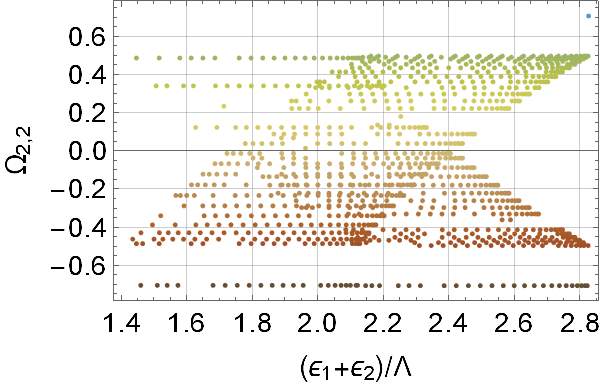}
    }
 }
 \raisebox{-1\height}{
    \parbox{\figWidth}{(e) $s=0.99$\\
    \includegraphics[width = 1\linewidth]{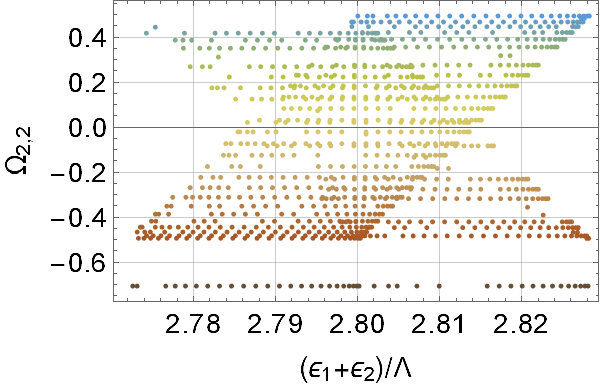}
    }
 }

\caption{\label{figDipoleMatrixElements2}Dipole matrix elements $\Omega_{2,2}$ between states with two quasi-particles excitations and energy $\epsilon_1$ and $\epsilon_2$. Shown are the projections onto the $(\epsilon_1+\epsilon_2)$ line. The $(\epsilon_1-\epsilon_2)$-dependence becomes irrelevant for long chains, as inferred by the analytic calculations. The gap closure in $\Omega_{2,2}$ shown for $s=0.75$ and $s=0.99$ marks an inversion of the band from particle to hole character, a key feature of a gapped topological phase, see analytical calculations and Sec.~\ref{sec:IdentificationOfTheTopologicalPhase}.}
\end{figure}

\begin{figure}
 \centering
 \raisebox{-1\height}{
    \parbox{\figWidth}{(a1) $s=0.01$\\
    \includegraphics[width = 1\linewidth]{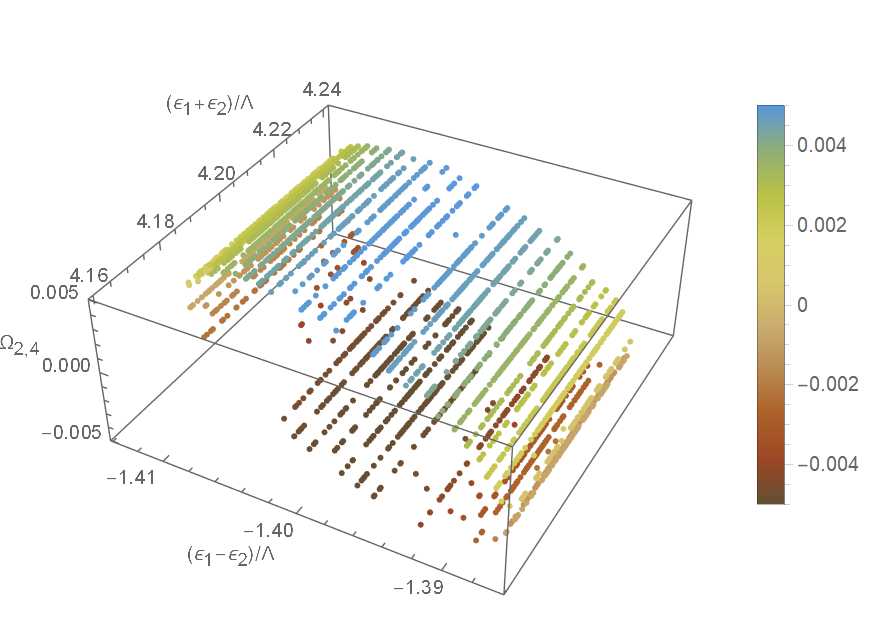}
    }
 }
 \raisebox{-1\height}{
    \parbox{\figWidth}{(b1) \\
    \includegraphics[width = 1\linewidth]{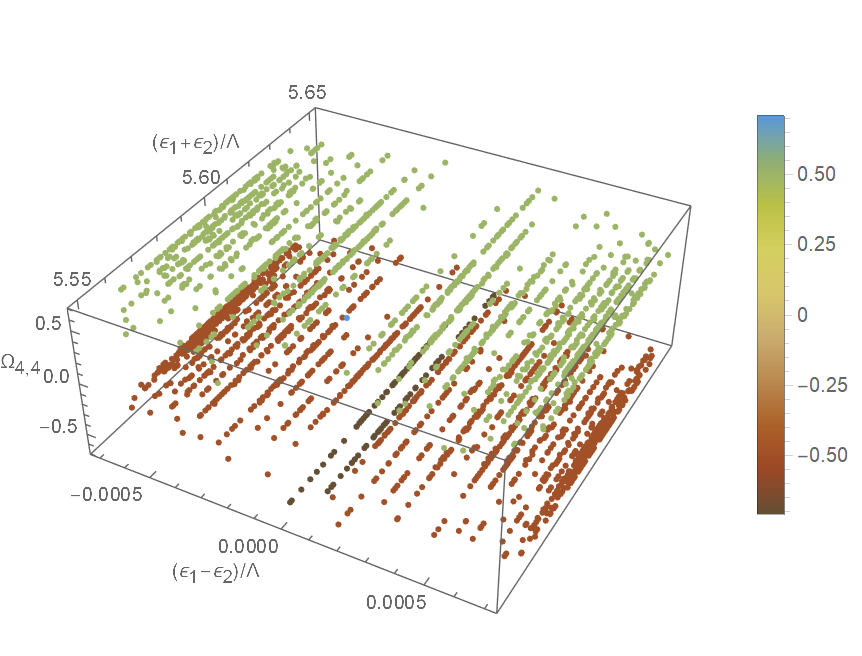}
    }
 }
 \\
 \raisebox{-1\height}{
    \parbox{\figWidth}{(a2) $s=0.25$\\
    \includegraphics[width = 1\linewidth]{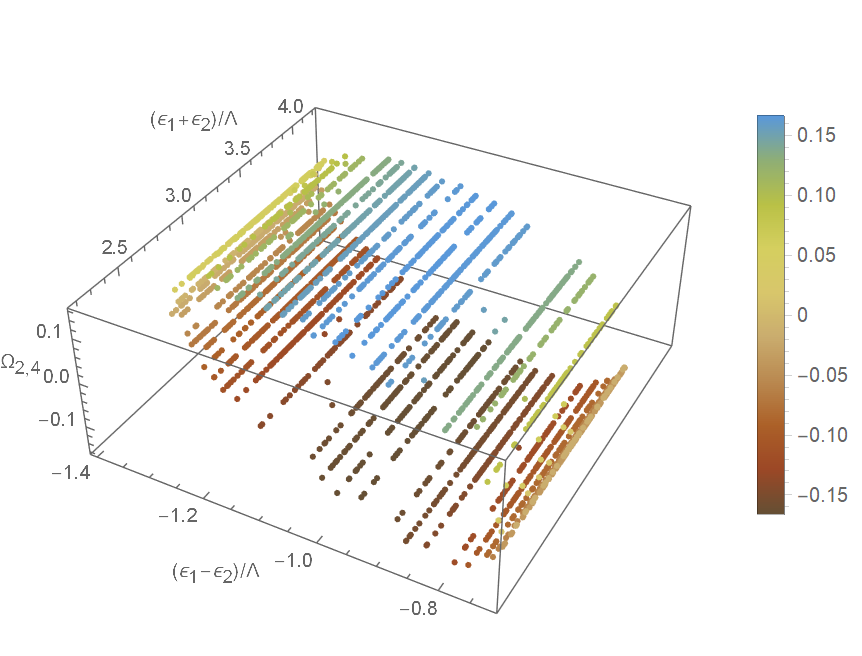}
    }
 }
 \raisebox{-1\height}{
    \parbox{\figWidth}{(b2)\\
    \includegraphics[width = 1\linewidth]{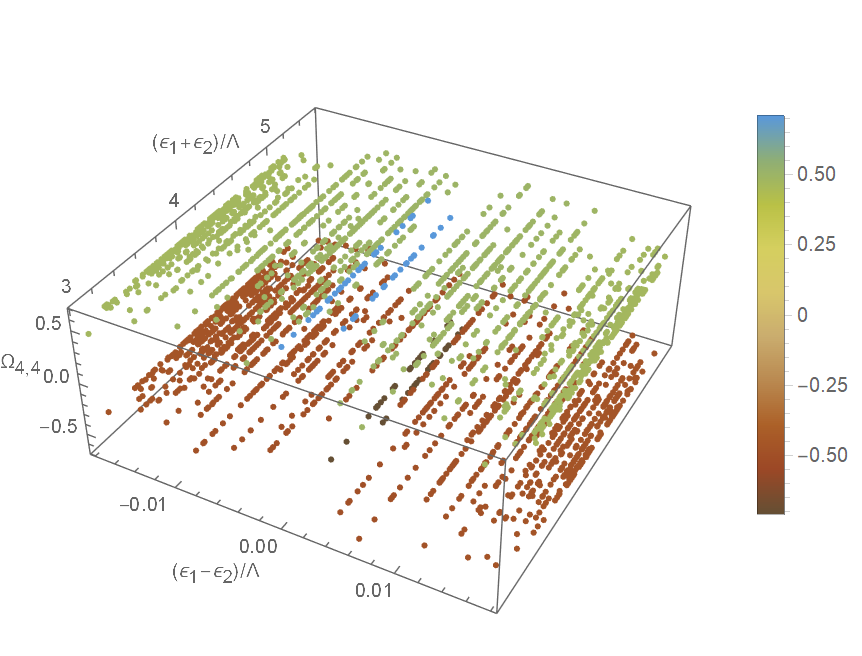}
    }
 }
 \\
 \raisebox{-1\height}{
    \parbox{\figWidth}{(a3)  $s=0.5$\\
    \includegraphics[width = 1\linewidth]{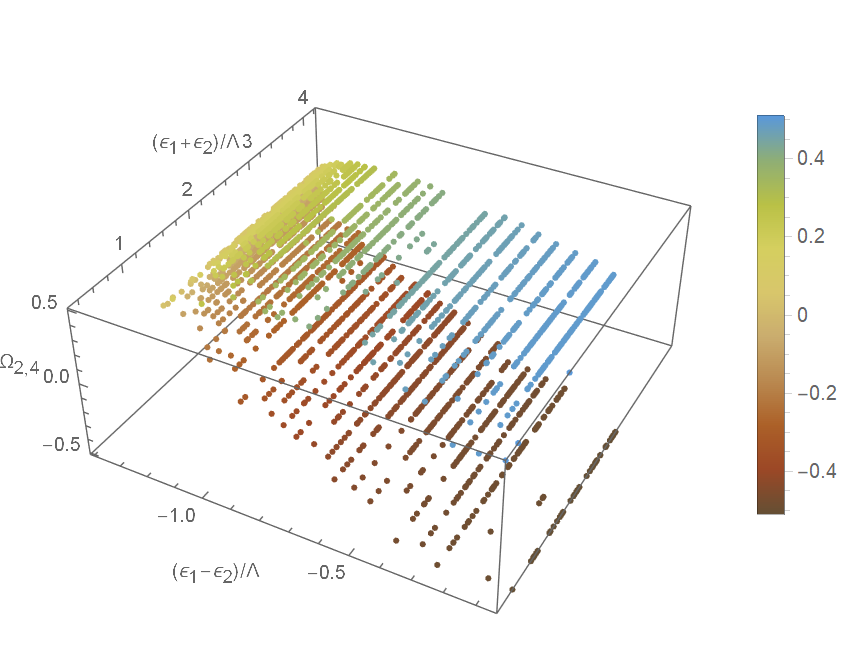}
    }
 }
 \raisebox{-1\height}{
    \parbox{\figWidth}{(b3)\\
    \includegraphics[width = 1\linewidth]{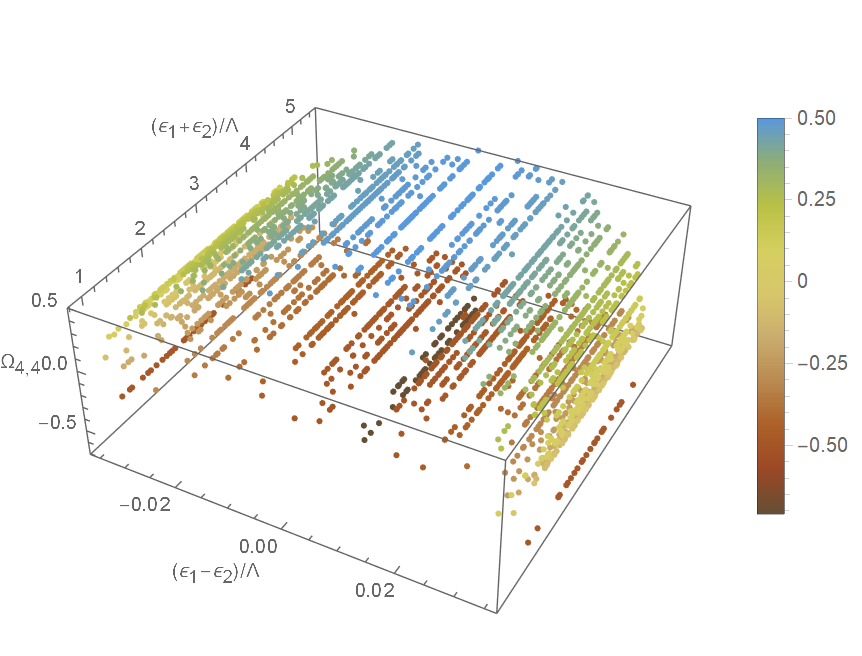}
    }
 }
 \\
 \raisebox{-1\height}{
    \parbox{\figWidth}{(a4)  $s=0.75$\\
    \includegraphics[width = 1\linewidth]{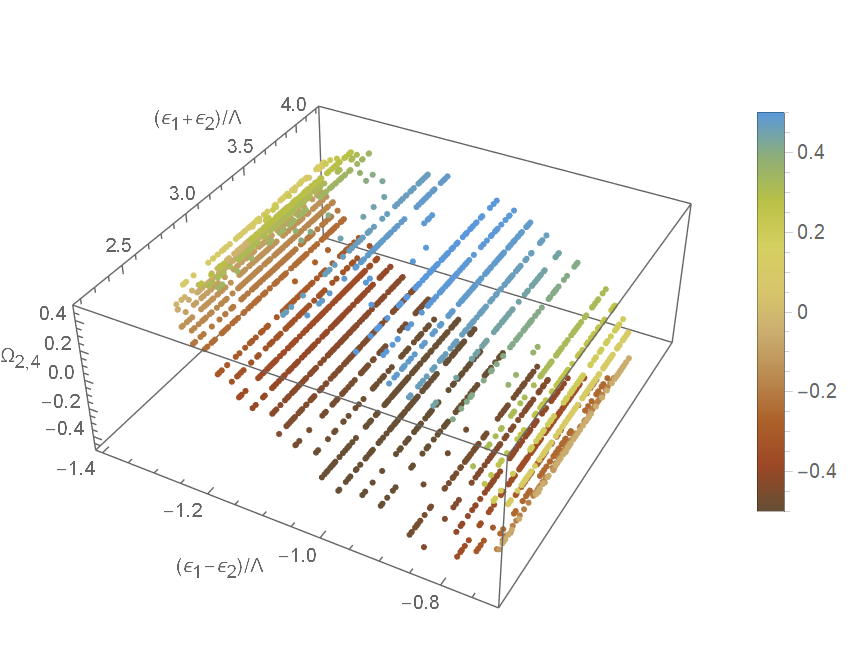}
    }
 }
 \raisebox{-1\height}{
    \parbox{\figWidth}{(b4)\\
    \includegraphics[width = 1\linewidth]{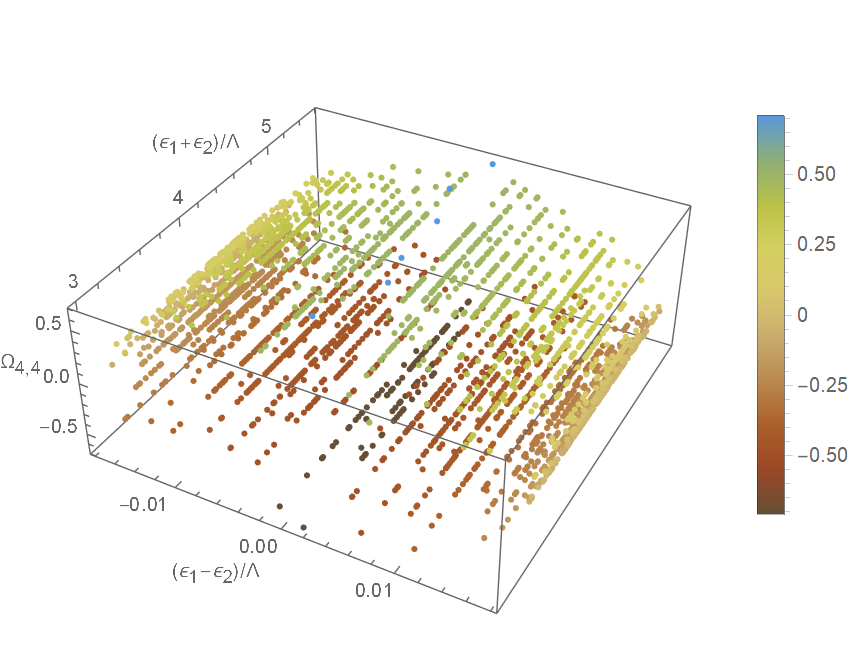}
    }
 }
 \\
 \raisebox{-1\height}{
    \parbox{\figWidth}{(a5)  $s=0.99$\\
    \includegraphics[width = 1\linewidth]{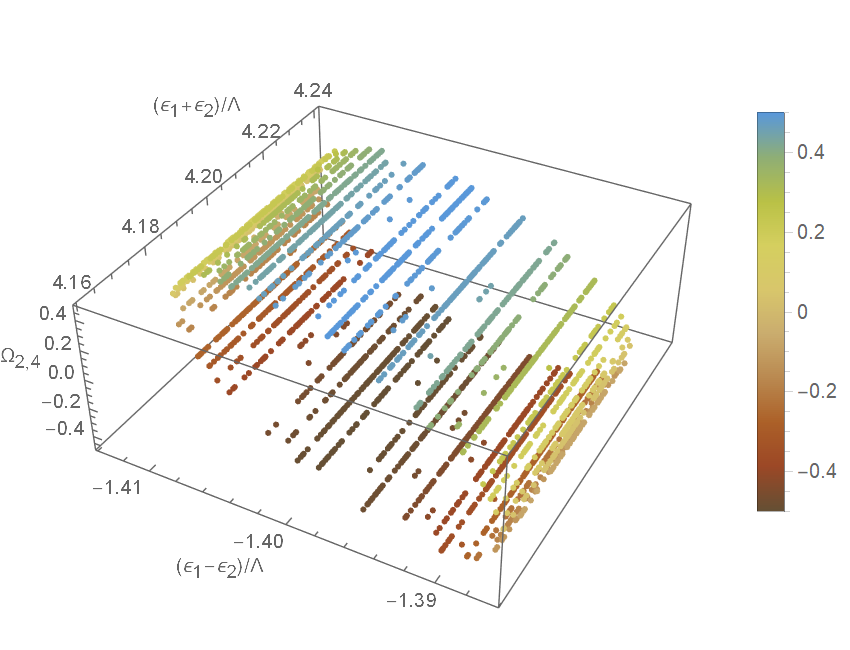}
    }
 }
 \raisebox{-1\height}{
    \parbox{\figWidth}{(b5)\\
    \includegraphics[width = 1\linewidth]{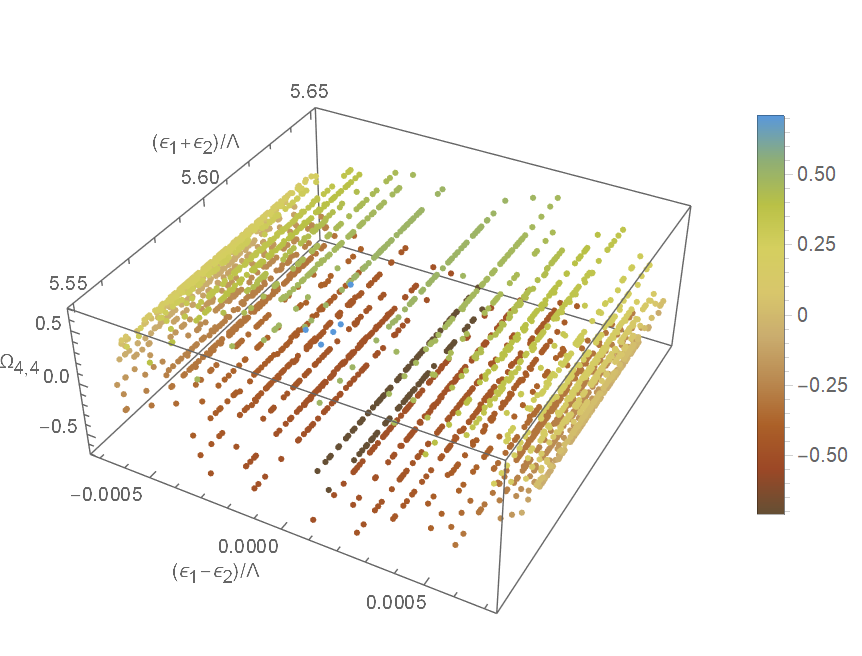}
    }
 }
\caption{\label{figDipoleMatrixElements4}The dipole matrix elements $\Omega_{a,b}$ between states with two and four quasiparticle excitations. Each diagram includes $100,000$ uniformly randomly drawn pairs of states at energies $\epsilon_1$ and $\epsilon_2$ (measured as differences from the ground state energy) and their corresponding matrix elements. The parameter $s$ takes on the values $0.01$, $0.25$, $0.5$, $0.75$, and $0.99$ from top to bottom.}
\end{figure}

We further evaluate the matrix elements of the $x$-component of the dipole operator numerically for two- and four-particles cases. Between these states, the vast majority of matrix elements vanishes because the dipole operator either changes the number of quasiparticles by $\pm2$ or leaves the number of quasiparticles unchanged. 
Further matrix elements vanish because the dipole operator only combines momentum modes that are close-by. This is seen by expressing the dipole operator in momentum space, i.e.,
\begin{align}
 d_x = -e R \sum_j \cos\left(\frac{2\pi}{N}j \right) a_j^\dagger a_j = -\frac{e R}{2} \sum_k \tilde{a}_{k+1}^\dagger \tilde{a}_k +h.c.,
\end{align}
with the Fourier transformed quasiparticle operator $\tilde{a}_{k}$.
The results are shown in Figs.~(\ref{figDipoleMatrixElements}-\ref{figDipoleMatrixElements4}), which depict the nonvanishing matrix elements $\Omega_{a,b}$ between states with $a$ quasiparticle excitations and $b$ quasiparticle excitations, where $a$ and $b$ are $0$, $2$, or $4$. 
The matrix elements connecting the groundstate with two-particle states as well as the matrix elements that connect two-particle states to two-particle states represent the energetically lowest states where signatures of braiding of quasiparticles can occur.
In fact, from $\Omega_{2,2}$, i.e., the dipole transitions between states with two quasiparticles, we observe a gap for the topologically trivial phase at $s<0.5$, which is closed in the topologically nontrivial phase $s>0.5$. As we elaborate in Secs.~\ref{sec:2PTo2P} and~\ref{sec:IdentificationOfTheTopologicalPhase}, the closure of the gap in $\Omega_{2,2}$ indicates a change of the character of an electronic band from particle-like to hole-like. This is a key feature of gapped topological phases of matter.

\section{\label{sec:analytic}Analytic Details and Discussion}

\subsection{Eigenmodes and Eigenenergies}

\noindent
First, we briefly summarize the solution of the Kitaev chain as given by Kitaev in his original paper from 2001, see Ref.~\cite{Kitaev2001}. The Hamiltonian of the periodic Kitaev chain is
\begin{equation}
\begin{split}
   \label{eq:KitaevHamiltonianS}
    H = \sum_{n=1}^N \left[ -w a_{n+1}^\dagger a_n - \mu a_n^\dagger a_n + \Delta a_n a_{n+1} \right] + \text{h.c.},
\end{split}
\end{equation}
with the hopping amplitude $w$, the chemical potential $2\mu$, and the complex superconducting gap parameter $\Delta = \abs{\Delta} e^{i\varphi}$. Periodic boundary conditions are enforced by $a_{N+1} \equiv a_1$. The system is gauge invariant under the change of the superconducting phase $\varphi$. The Hamiltonian is diagonalized by the eigenmodes (Bogoliubons)
\begin{equation}
\label{eq:bogoliubons}
f_k = 
\begin{cases}
    e^{i(\frac{\pi}{4} + \frac{\varphi}{2})} \cos(\frac{\vartheta_k}{2}) \tilde{a}_k + e^{-i(\frac{\pi}{4} + \frac{\varphi}{2})} \sin(\frac{\vartheta_k}{2}) \tilde{a}_{-k}^\dagger, & \text{if } w\cos\left(\frac{2\pi}{N}k\right) + \mu > 0, \\
    - e^{i(\frac{\pi}{4} + \frac{\varphi}{2})} \sin(\frac{\vartheta_k}{2}) \tilde{a}_{k} + e^{-i(\frac{\pi}{4} + \frac{\varphi}{2})} \cos(\frac{\vartheta_k}{2}) \tilde{a}_{-k}^\dagger, & \text{if } w\cos\left(\frac{2\pi}{N}k\right) + \mu < 0,
\end{cases}
\end{equation}
with the Fourier transformed operators $\tilde{a}_k = (1 /\sqrt{N}) \sum_{j=1}^N \exp\left(2\pi i j k/N\right) a_j$ and the mixing angle
\begin{equation}
    \vartheta_k = \arctan(\frac{\abs{\Delta}\sin(\frac{2\pi}{N}k)}{w\cos(\frac{2\pi}{N}k)+\mu}).
\end{equation}
The Bogoliubov transformation mixes particle and hole states. We can choose the Bogoliubov transformation to be orthogonal by gauge fixing $\varphi=-\pi/4$.
The diagonalized Hamiltonian reads
\begin{equation}
    H = \sum_{k=1}^N \epsilon_k \left(f_k^\dagger f_k - \frac{1}{2}\right) = \sum_{k=1}^N  \frac{\epsilon_k}{2} \left( f_k^\dagger f_k - f_k f_k^\dagger\right),
\end{equation}
with the dispersion relation \cite{Kitaev2001}
\begin{equation}
\label{eq:dispersion}
    \epsilon_k = 2 \sqrt{\left(w\cos(\frac{2\pi}{N}k) + \mu\right)^2 + \abs{\Delta}^2 \sin^2\left(\frac{2\pi}{N}k\right)}.
\end{equation}
Note that for $\abs{w}<\abs{\mu}$, the eigenmodes have either predominantly particle or predominantly hole character depending on the sign of the chemical potential. For $\abs{w}>\abs{\mu}$ and $\Delta\neq 0$, there are distinct $k$-regions with particle and with hole character. The change from particle to hole character happens at the inversion points $\frac{2\pi}{N}k_\text{inv} = \arccos\left(\frac{-\mu}{w}\right)$ and $\frac{2\pi}{N} (-k_\text{inv}) \equiv 2\pi - \frac{2\pi}{N}k_\text{inv}$. These relations distinguish two different gapped electronic topological phases. The first is called the trivial and the latter the non-trivial phase. At the boundary in parameter space, $w=\mu$, the band gap closes. In fact, one cannot interpolate between the two phases without closing the band gap. This leads to zero-energy modes at domain walls between different phases as well as at the ends of an open non-trivial Kitaev chain. These zero-modes are conjectured to have Majorana character \cite{Kitaev2001}. For the following discussion, we fix the maximal excitation energy of a single Bogoliubon
\begin{equation}
\label{eq:lambda}
    \Lambda := \max(\epsilon_k),
\end{equation}
as in the main text.

\subsection{\label{sec:dipoleOperator}The Dipole Operator}

\noindent
The dipole operator is the sum of the positions of each site times the charge on the same site. For the ring geometry, its $x$-component reads
\begin{align}
\begin{split}
\label{eq:analDipole}
 d_x &= -e R \sum_j \cos(\frac{2\pi}{N}j ) a_j^\dagger a_j = - \frac{eR}{2} \sum_k \tilde{a}_{k+1}^\dagger \tilde{a}_k +h.c. \\
 &= -e \frac{R}{2} \sum_k \operatorname{sgn}_k \bigg[\left(c_k c_{k+1} - s_k s_{k+1}\right)  \left(f_{k+1}^\dagger f_{k} + f_{k}^\dagger f_{k+1}\right)  + s_k c_{k+1} \left( f_{k+1}^\dagger f_{-k}^\dagger - f_{k+1} f_{-k} \right)  \\
 & \hspace{2.8cm} + s_k c_{k-1} \left( f_{k-1}^\dagger f_{-k}^\dagger - f_{k-1} f_{-k} \right) \bigg],
\end{split}
\end{align}
where $c_k = \cos(\frac{\vartheta_k}{2})$, $s_k = \sin(\frac{\vartheta_k}{2})$ and $\operatorname{sgn}_k = \operatorname{sgn}\left(w \cos\left(\frac{2\pi}{N} k\right) + \mu\right)$.
Note that Eq.~(\ref{eq:analDipole}) does not directly follow from Eq.~(\ref{eq:bogoliubons}) for $\abs{w}>\abs{\mu}$ (non-trivial phase). We would rather need to replace the terms that cross the inversion points. This leads to new terms
\begin{align}
\begin{split}
eR \bigg[ &\left(s_{\floor{k_\text{inv}}} c_{\floor{k_\text{inv}}+1} + c_{\floor{k_\text{inv}}} s_{\floor{k_\text{inv}}+1} \right) f_{\floor{k_\text{inv}}+1}^\dagger f_{\floor{k_\text{inv}}} - \left(s_{\floor{k_\text{inv}}} c_{\floor{k_\text{inv}}+1} + c_{\floor{k_\text{inv}}} s_{\floor{k_\text{inv}}+1} \right) f_{\floor{-k_\text{inv}}}^\dagger f_{\floor{-k_\text{inv}}-1} \\ &+ \left(c_{\floor{k_\text{inv}}} c_{\floor{k_\text{inv}}+1} - s_{\floor{k_\text{inv}}} s_{\floor{k_\text{inv}}+1}\right) \left( f_{\floor{k_\text{inv}}} f_{\floor{-k_\text{inv}}-1} + f_{\floor{k_\text{inv}}+1}^\dagger f_{\floor{-k_\text{inv}}}^\dagger \right) \bigg],
\end{split}
\end{align}
where $\floor{\cdot}$ is the floor function giving the largest integer smaller than or equal to the argument. We will not consider this technical term any further because for large $N$ this term only has an influence on a null set in $k$-space.

We go to the interaction picture by replacing $f_j \mapsto f_j (t) = e^{-i t \epsilon_j} f_j$.
The time-dependent dipole operator reads
\begin{align}
\begin{split}
 d_x (t) = -e \frac{R}{2} \sum_k \operatorname{sgn}_k \bigg[&\left(c_k c_{k+1} - s_k s_{k+1}\right)  \left( e^{i t \left(\epsilon_{k+1} - \epsilon_{k}\right)}  f_{k+1}^\dagger f_{k} + e^{i t \left(\epsilon_{k} - \epsilon_{k+1}\right)} f_{k}^\dagger f_{k+1}\right) \\
 &+ s_k c_{k+1} \left( e^{i t \left(\epsilon_{k+1} + \epsilon_{k} \right)} f_{k+1}^\dagger f_{-k}^\dagger - e^{-i t \left(\epsilon_{k+1} + \epsilon_{k}\right)} f_{k+1} f_{-k} \right) \\
 &+ s_k c_{k-1} \left( e^{i t \left(\epsilon_{k-1} + \epsilon_{k}\right)} f_{k-1}^\dagger f_{-k}^\dagger - e^{-i t \left(\epsilon_{k-1} + \epsilon_{k}\right)} f_{k-1} f_{-k} \right) \bigg].
\end{split}
\end{align}
Notice that for large $N$, the coefficients converge to the following functions that can be expressed in simple terms by the system parameters:
\begin{align}
\label{eq:transitionpm2}
\operatorname{sgn}_k \left(s_k c_{k+1}\right) e^{i t \left(\epsilon_{k+1} + \epsilon_{k}\right)} &\xrightarrow[N\to\infty]{} \frac{\abs{\Delta}\sin\left(\frac{2\pi}{N}k\right)}{\epsilon_k} e^{i 2\epsilon_{k} t}, \\
\label{eq:transitionpm0}
\operatorname{sgn}_k \left(c_k c_{k+1} - s_k s_{k+1}\right) e^{i t \left(\epsilon_{k+1} - \epsilon_{k}\right)} &\xrightarrow[N\to\infty]{} \frac{2w\cos\left(\frac{2\pi}{N}k\right) + 2\mu}{\epsilon_k}.
\end{align}
The first coefficient corresponds to increasing or decreasing the number of quasiparticles by two. It is the same (up to a scaling factor $\Delta$) for Hamiltonians with the same band structure. In particular, it will be the same if the topological phases differ, but the band structures coincide, which can happen, see the main text. The second coefficient corresponds to transitions in the same quasiparticle sector. Here, it can be seen that the transition dipole moment has zeroes at the inversion points if and only if the system is in the topologically non-trivial phase. Also notice the different values of the transition energies. For the transitions between quasiparticle sectors, the transition energy is $2 E_\text{Gap} \leq \epsilon_{k+1} + \epsilon_{k}\leq 2\Lambda$. For transitions within a quasiparticle sector, the transition energy is $\abs{\epsilon_{k+1} - \epsilon_{k}} \leq \Lambda - E_\text{Gap}$, and goes to zero for large $N$.
We find that the latter contributions appear in the 2D spectra in the main text as the horizontal peaks. These peaks will have a zero at the energies of the inversion points if and only if the system is in the topologically non-trivial phase. For intermediate and large $\Delta$, this zero splits the peak continuum along the horizontal axis in two parts, as can be seen in the main text. For small $\Delta$, the inversion points are at the lower band edge, so the zero appears at the small frequency end of the horizontal peak continuum. This zero can be identified by comparing with an absorption spectrum or the peaks on the diagonal of the 2D spectrum. 

\subsection{Matrix Elements}

\noindent
Here, we provide explicit expressions for the matrix elements for the transitions from the ground state to the two-quasiparticle sector and for transitions within the two-quasiparticle sector.

We denote the quasiparticle vacuum and ground state by $\ket{\Omega}$ and use the convention $\ket{k_1, \dots, k_n} := f_{k_n} \cdot\dots\cdot f_{k_1} \ket{\Omega}$ for $k_1 \leq \dots \leq k_n$.

\subsubsection{Groundstate}
\noindent
The groundstate's dipole moment vanishes, i.e., 
\begin{equation}
\bra{\Omega} d_x (t) \ket{\Omega} = 0 .
\end{equation}

\subsubsection{Groundstate to 2-particle-States}
\label{sctnGroundStateTo2ParticleState}
\noindent
For the transitions from the groundstate to the two-quasiparticle sector, the transition dipole moments are of the form
\begin{align}
\begin{split}
&-\frac{2}{eR}\bra{k_1, k_2} d_x (t) \ket{\Omega} \\
= &\sum_k \operatorname{sgn}_k s_k c_{k+1} e^{i t \left(\epsilon_{k+1} + \epsilon_{k} \right)} \bra{\Omega} f_{k_1} f_{k_2} f_{k+1}^\dagger f_{-k}^\dagger \ket{\Omega} + \sum_k \operatorname{sgn}_k s_k c_{k-1} e^{i t \left(\epsilon_{k-1} + \epsilon_{k} \right)} \bra{\Omega} f_{k_1} f_{k_2} f_{k-1}^\dagger f_{-k}^\dagger \ket{\Omega}.
\end{split}
\end{align}
The vacuum expectation values are evaluated by Wick contractions, i.e., 
\begin{align}
\begin{split}
\bra{\Omega} f_{k_1} f_{k_2} f_{k+1}^\dagger f_{-k}^\dagger \ket{\Omega} &= \bra{\Omega} \wick{ \c1 f_{k_1} \c2 f_{k_2} \c1 f_{k+1}^\dagger \c2 f_{-k}^\dagger} \ket{\Omega} + \bra{\Omega} \wick{ \c2 f_{k_1} \c1 f_{k_2} \c1 f_{k+1}^\dagger \c2 f_{-k}^\dagger} \ket{\Omega} \\ 
&= -\bra{\Omega} \wick{ \c1 f_{k_1} \c1 f_{k+1}^\dagger \c1 f_{k_2}  \c1 f_{-k}^\dagger} \ket{\Omega} + \bra{\Omega} \wick{ \c1 f_{k_2} \c1 f_{k+1}^\dagger \c1 f_{k_1}  \c1 f_{-k}^\dagger} \ket{\Omega} \\
&= - \delta_{k_1, k+1} \delta_{k_2, -k} + \delta_{k_2, k+1} \delta_{k_1, -k} \, ,
\end{split}
\\
\bra{\Omega} f_{k_1} f_{k_2} f_{k-1}^\dagger f_{-k}^\dagger \ket{\Omega} &= - \delta_{k_1, k-1} \delta_{k_2, -k} + \delta_{k_2, k-1} \delta_{k_1, -k}.
\end{align}
As a result, we obtain
\begin{equation}
\bra{k_1, k_2} d_x (t) \ket{\Omega} = 
\begin{cases}
    \frac{eR}{2} \operatorname{sgn}_{k_1} \left(s_{k_1-1} c_{k_1} + s_{k_1} c_{k_1-1} \right) e^{i t \left(\epsilon_{k_1} + \epsilon_{k_1-1} \right)}, & \text{if } k_1 + k_2 = 1,\\
    \frac{eR}{2} \operatorname{sgn}_{k_1} \left(s_{k_1+1} c_{k_1} + s_{k_1} c_{k_1+1} \right) e^{i t \left(\epsilon_{k_1+1} + \epsilon_{k_1} \right)}, & \text{if } k_1 + k_2 = -1,\\
    0,               & \text{otherwise.}
\end{cases}
\end{equation}
For large $N$, the only non-vanishing matrix element is
\begin{equation}
\label{eq:dipole0to2}
    \bra{k\pm dk, -k} d_x (t) \ket{\Omega} = \operatorname{sgn}_{k} \frac{eR x_k}{2\sqrt{x_k^2 + 1}} e^{i 2\epsilon_k t} = \frac{eR\abs{\Delta}\sin\left(\frac{2\pi}{N}k\right)}{\epsilon_k} e^{i 2\epsilon_k t},
\end{equation}
where $dk$ is an infinitesimal shift in momentum space and
\begin{equation}
    x_k = \frac{\abs{\Delta}\sin(\frac{2\pi}{N}k)}{w\cos(\frac{2\pi}{N}k)+\mu}.
\end{equation}
In linear spectroscopy, only the absolute squared of the dipole moment in Eq.~(\ref{eq:dipole0to2}) enters. The form of the absorption spectra are fully determined by the dispersion relation $\epsilon_k$. If the band structures of two Hamiltonians coincide, they will give rise to the same absorption spectrum modulo a scaling factor of $\abs{\Delta}^2$. Hence, the different topological phases are indistinguishable by linear spectroscopy methods. 

\subsubsection{\label{sec:2PTo2P}2-particle-States to 2-particle-States}
\noindent
Similar as before, the 2-particle to 2-particle transition dipole moments are of the form
\begin{align}
\begin{split}
&-\frac{2}{eR}\bra{k_1, k_2} d_x (t) \ket{l_1, l_2} \\
= &\sum_k \left(c_k c_{k+1} - s_k s_{k+1}\right)  \left( e^{i t \left(\epsilon_{k+1} - \epsilon_{k}\right)} \bra{\Omega} f_{k_1} f_{k_2} f_{k+1}^\dagger f_{k} f_{l_2}^\dagger f_{l_1}^\dagger \ket{\Omega}
+ e^{i t \left(\epsilon_{k} - \epsilon_{k+1}\right)} \bra{\Omega} f_{k_1} f_{k_2} f_{k}^\dagger f_{k+1} f_{l_2}^\dagger f_{l_1}^\dagger \ket{\Omega} \right),
\end{split}
\end{align}
with the vacuum expectation values
\begin{align}
\begin{split}
\bra{\Omega} f_{k_1} f_{k_2} f_{k+1}^\dagger f_{k} f_{l_2}^\dagger f_{l_1}^\dagger \ket{\Omega} =
&\bra{\Omega} \wick{ \c1 f_{k_1} \c2 f_{k_2} \c1 f_{k+1}^\dagger \c3 f_{k} \c2 f_{l_2}^\dagger \c3 f_{l_1}^\dagger } \ket{\Omega} 
+ \bra{\Omega} \wick{ \c1 f_{k_1} \c2 f_{k_2} \c1 f_{k+1}^\dagger \c1 f_{k} \c1 f_{l_2}^\dagger \c2 f_{l_1}^\dagger } \ket{\Omega} \\
+ &\bra{\Omega} \wick{ \c1 f_{k_1} \c2 f_{k_2} \c2 f_{k+1}^\dagger \c2 f_{k} \c1 f_{l_2}^\dagger \c2 f_{l_1}^\dagger } \ket{\Omega}
+ \bra{\Omega} \wick{ \c1 f_{k_1} \c2 f_{k_2} \c2 f_{k+1}^\dagger \c2 f_{k} \c2 f_{l_2}^\dagger \c1 f_{l_1}^\dagger } \ket{\Omega} \\
= &\delta_{k_1, k+1}\delta_{k_2, l_2}\delta_{k, l_1} - \delta_{k_1, k+1}\delta_{k_2, l_1}\delta_{k, l_2} - \delta_{k_1, l_2}\delta_{k_2, k+1}\delta_{k, l_1} + \delta_{k_1, l_1}\delta_{k_2, k+1}\delta_{k, l_2},
\end{split} \\
\begin{split}
\bra{\Omega} f_{k_1} f_{k_2} f_{k}^\dagger f_{k+1} f_{l_2}^\dagger f_{l_1}^\dagger \ket{\Omega} =
&\bra{\Omega} \wick{ \c1 f_{k_1} \c2 f_{k_2} \c1 f_{k}^\dagger \c3 f_{k+1} \c2 f_{l_2}^\dagger \c3 f_{l_1}^\dagger } \ket{\Omega} 
+ \bra{\Omega} \wick{ \c1 f_{k_1} \c2 f_{k_2} \c1 f_{k}^\dagger \c1 f_{k+1} \c1 f_{l_2}^\dagger \c2 f_{l_1}^\dagger } \ket{\Omega} \\
+ &\bra{\Omega} \wick{ \c1 f_{k_1} \c2 f_{k_2} \c2 f_{k}^\dagger \c2 f_{k+1} \c1 f_{l_2}^\dagger \c2 f_{l_1}^\dagger } \ket{\Omega}
+ \bra{\Omega} \wick{ \c1 f_{k_1} \c2 f_{k_2} \c2 f_{k}^\dagger \c2 f_{k+1} \c2 f_{l_2}^\dagger \c1 f_{l_1}^\dagger } \ket{\Omega} \\
= &\delta_{k_1, k}\delta_{k_2, l_2}\delta_{k+1, l_1} - \delta_{k_1, k}\delta_{k_2, l_1}\delta_{k+1, l_2} - \delta_{k_1, l_2}\delta_{k_2, k}\delta_{k+1, l_1} + \delta_{k_1, l_1}\delta_{k_2, k}\delta_{k+1, l_2}.
\end{split}
\end{align}
For large $N$, it is a good and convenient assumption that $-k \ll k \pm 1$ almost everywhere.
Then, the only transition dipole moments relevant for the 2D spectroscopy are
\begin{align}
\bra{k_1, k_2} d_x (t) \ket{-k, k\pm1} = -\frac{eR}{2}
\begin{cases}
\left(c_{k_1-1} c_{k_1} - s_{k_1-1} s_{k_1}\right) e^{i t \left(\epsilon_{k_1} - \epsilon_{k_1-1}\right)}
, & \text{if } k_1 = -k+1 \ \text{and } k_2 = k\pm1,\\
\left(c_{k_2-1} c_{k_2} - s_{k_2-1} s_{k_2}\right) e^{i t \left(\epsilon_{k_2} - \epsilon_{k_2-1}\right)}
, & \text{if } k_1 = -k \ \text{and } k_2 = k+1\pm1,\\
\left(c_{k_1} c_{k_1+1} - s_{k_1} s_{k_1+1}\right) e^{i t \left(\epsilon_{k_1} - \epsilon_{k_1+1}\right)}
, & \text{if } k_1 = -k-1 \ \text{and } k_2 = k\pm1,\\
\left(c_{k_2} c_{k_2+1} - s_{k_2} s_{k_2+1}\right) e^{i t \left(\epsilon_{k_2} - \epsilon_{k_2+1}\right)}
, & \text{if } k_1 = -k \ \text{and } k_2 = k-1\pm1.
\end{cases}
\end{align}
For large $N$, the discussion of the dipole moments is analogous to the one at the end of subsection \ref{sec:dipoleOperator}.

\subsubsection{2-particle States to 4-particle States}
\noindent
In a similar way, but with increasing combinatorial effort, the transitions from the two-quasiparticle sector to the four-quasiparticle sector can be obtained. Their effect on the 2D spectra will be similar to what is already discussed at the end of subsection \ref{sec:dipoleOperator}. Representative numerical values are shown in Fig.~\ref{figDipoleMatrixElements4}.

\section{\label{sec:IdentificationOfTheTopologicalPhase}Identification of the Topological Phase}

\noindent
Below, we summarize our findings in Secs.~\ref{sec:matrixElements}~and~\ref{sec:analytic} that are relevant for the interpretation of the 2D spectra shown in the main text. The only contributions to the four-point correlation functions in Eqs.~(\ref{eq:C1}) to (\ref{eq:C4}) come from transitions between the groundstate and two-particle states, from transitions within the two-particle sector, and from transitions between the two-particle sector to the four-particle sector. As seen from Figs.~\ref{figDipoleMatrixElements} and \ref{figDipoleMatrixElements4}, the transition dipole moments from the ground state to two-particle states and the transition dipole moments from two-particle to four-particle states show no qualitative difference for Hamiltonians with the same dispersion relation. In fact, they merely differ by a scaling factor that is the ratio between the supeconducting gap parameters of the two Hamiltonians $\abs{\Delta^\prime} / \abs{\Delta}$ as derived in subsection~\ref{sec:dipoleOperator}. We observe in the numerical computations in Figs.~\ref{figDipoleMatrixElements} and \ref{figDipoleMatrixElements4} that the scaling factors are in agreement with $\abs{\Delta^\prime} / \abs{\Delta}=99$ for $s=0.01$ and $s^\prime=0.99$, and $\abs{\Delta^\prime} / \abs{\Delta}=3$ for $s=0.25$ and $s^\prime=0.75$. This scaling factor is fully determined by the condition that the band structures coincide and the topological phases differ as seen in Eq.~\ref{eq:invariantmap} of the main text. Yet, such a scaling factor is not a conclusive signature which  characterizes the topological phases in a unique way. Instead a qualitative difference in the 2D spectra of the topological phases stems from the transitions within the two-particle sector. For these transitions, we observe differences in Fig.~\ref{figDipoleMatrixElements}, in particular, a cluster of transitions with close-to-zero dipole moments in the non-trivial phase that is absent in the trivial phase. We characterize these zeroes by the analytic result in subsection~\ref{sec:dipoleOperator}. Eq.~(\ref{eq:transitionpm0}) and show that the two-to-two-particle transition dipole moments are always finite in the trivial phase but cross zero in the non-trivial phase. The crossing from negative to positive occurs precisely at the inversion points in $k$-space where the band changes from a predominantly particle to a predominantly hole band. Hence, these zeroes are a consequence of the band inversion happening in the non-trivial phase.
In the spectra, the transitions within the two-particle sector appear as low-frequency transitions while the transitions that change the number of quasiparticles must overcome at least twice the band gap. Therefore, the contributions that are characteristic for a topological phase should appear on the horizontal of the 2D spectra. There, they section the continuum in two precisely at the energy of the inversion points as seen in Fig.~\ref{fig:half_gapped}c in the main text. For small $\Delta$, i.e., $\abs{\Delta}^2 < w^2 - \mu^2$, this energy is found at the lower band edge and is only identifiable by comparison with linear spectra or the diagonal of the 2D spectrum.

\section{\label{sec:Rashba}Spectroscopy operator of the Rashba wire}

\noindent The Kitaev chain is the archetype of one-dimensional topological superconductors. However, it only appears in nature as a low-energy description of specifically engineered mesoscopic systems that require auxiliary effects such as proximity-induced superconductivity, spin-orbit coupling and strong magnetic fields. The general form of the spectroscopy operator given by the projection of the dipole operator onto the low-energy theory is a priori unclear. In this section, we demonstrate that the spectroscopy operator as given in the main text is the principal contribution to the spectroscopy operator of a semiconducting wire.

The Rashba wire is conjectured to realize topological superconductivity. It is a semiconducting wire on an $s$-wave superconducting substrate with strong spin-orbit coupling and Zeeman splitting due to applied external magnetic fields. It was first realized in an experiment reported in Ref.\  \cite{Kouwenhoven2012}. Following the exposition in Ref.~\cite{Alicea2011}, we derive its spectroscopy operator for the low-energy band in the limit of large magnetic fields $B$. The Hamiltonian of the Rashba wire is
\begin{equation}
   H = \int dx \left[  \begin{pmatrix} \psi^\dagger_{\uparrow x}, & \psi^\dagger_{\downarrow x} \end{pmatrix} \left( - \frac{\hbar^2 \partial^2_x}{2m} - \mu - i \hbar u (\mathbf{e} \cdot \boldsymbol{\sigma} ) \partial_x - \frac{g \mu_B B_z}{2} \sigma^z \right) \begin{pmatrix} \psi_{\uparrow x} \\ \psi_{\downarrow x} \end{pmatrix} + \left(\Delta \psi_{\downarrow x} \psi_{\uparrow x} + \text{h.c.} \right) \right],
\end{equation}
where $\mathbf{e} = \begin{pmatrix} e_x, & e_y, & 0 \end{pmatrix}$ with $e_x^2+e_y^2=1$. Further, $\mu$ is the chemical potential, $u$ the Dresselhaus/Rashba spin-orbit coupling strength and $\boldsymbol{\sigma}$ the vector of Pauli matrices.
The spin-singlet pairing $\Delta$ is due to proximity to an $s$-wave superconducting substrate.
For large magnetic fields, one can project the Hamiltonian onto a single-band model \cite{Alicea2011}. If $g \mu_B \abs{B_z} \gg \mu$ and $g \mu_B \abs{B_z} \gg \abs{\Delta}$, then $\psi_{\uparrow x} \sim \frac{\hbar u (e_x + i e_y)}{g \mu_B \abs{B_z}} \partial_x \Psi_x$ and $\psi_{\downarrow x} \sim \Psi_x$. The low-energy Hamiltonian is
\begin{equation}
   H \sim \int dx \left[ \Psi^\dagger_x \left( - \frac{\hbar^2 \partial^2_x}{2m} - \mu_\text{eff}\right) \Psi_x + \left(\Delta_\text{eff} \Psi_{x} \partial_x \Psi_{ x} + \text{h.c.} \right) \right] \, ,
\end{equation}
with $\mu_\text{eff} = \mu + g \mu_B \abs{B_z} / 2$ and
\begin{equation}
   \Delta_\text{eff} \approx \frac{\hbar u \Delta}{g \mu_B \abs{B_z}}  (e_x + i e_y).
\end{equation}
We proceed to express the dipole operator in terms of the field operator $\Psi_x$.
The dipole operator is 
\begin{equation}
   \mathbf{d} = - e \int dx \, \mathbf{r}_x \left( \psi^\dagger_{\downarrow x} \psi_{\downarrow x} + \psi^\dagger_{\uparrow x} \psi_{\uparrow x} \right),
\end{equation}
where $\mathbf{r}_x$ denotes the physical position at $x$. Here, we consider $x$ to merely parametrize the one-dimensional chain and not to distinguish any spatial direction.
Projecting the dipole operator onto the low-energy band yields
\begin{equation}
   \mathbf{d} \sim - e \int dx \,  \mathbf{r}_x \Psi^\dagger_{x} \Psi_{x} - \frac{2emu^2}{\hbar^2g^2\mu_B^2\abs{B_z}^2} \int dx \,  \mathbf{r}_x \Psi^\dagger_{x} \left( - \frac{\hbar^2\partial_x^2}{2m} \right) \Psi_{x}.
\end{equation}
We can already see here that the principal term is of the same kind as for the dipole operator given in the main text and the additional hopping term is negligible for sufficiently large $B$-fields. 

\begin{figure*}
\subfloat{\includegraphics[scale=0.5]{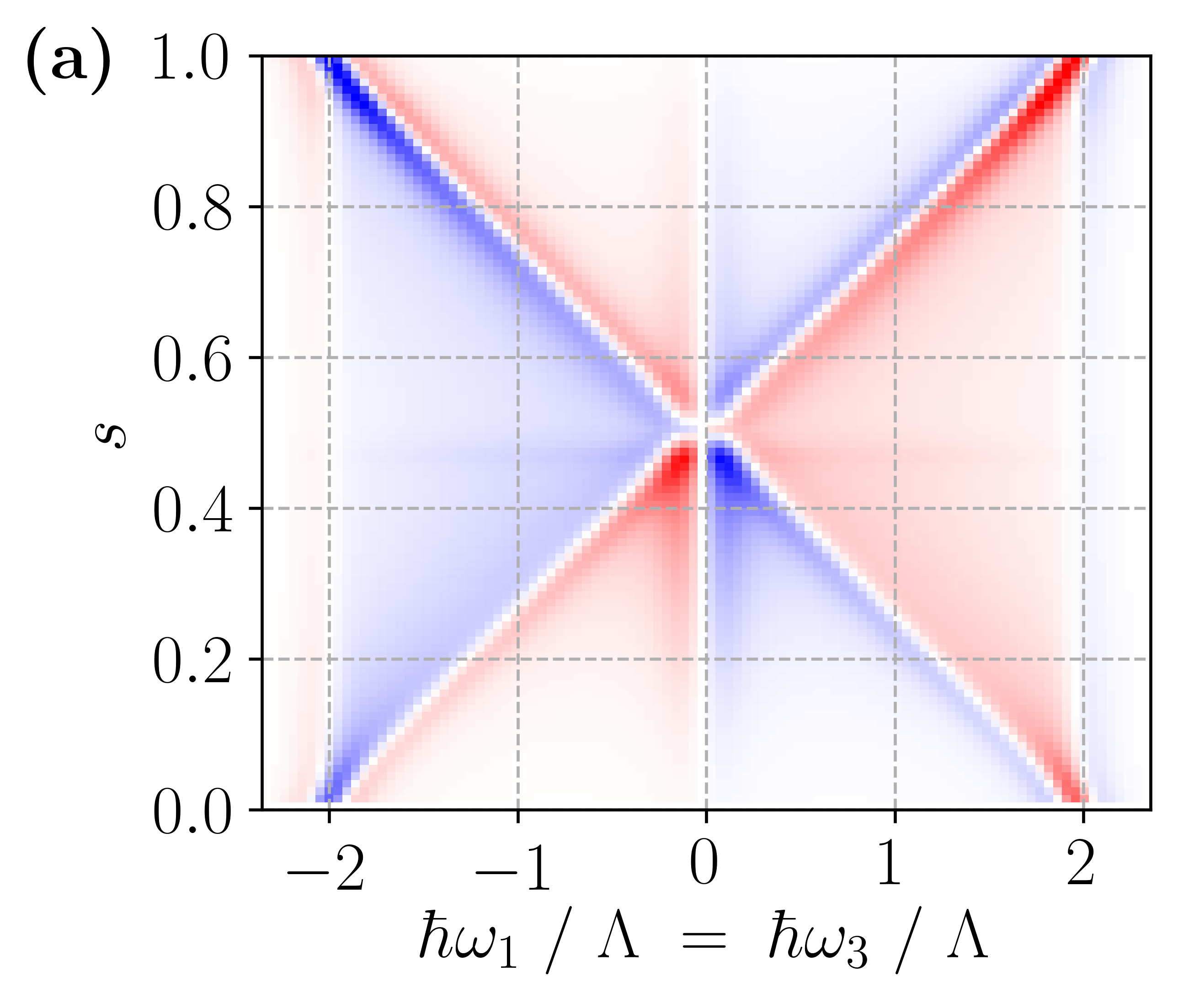}}
\subfloat{\includegraphics[scale=0.5]{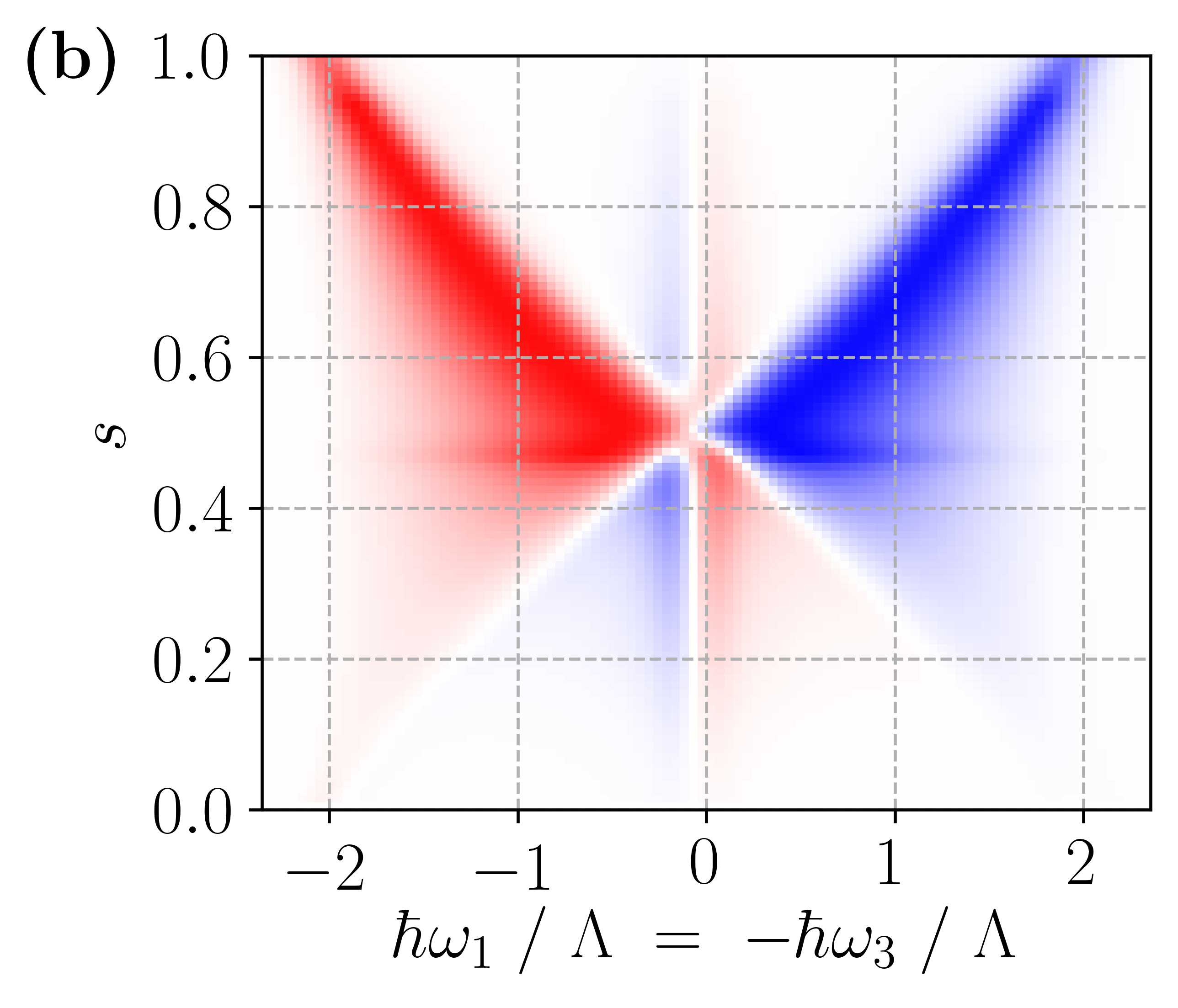}}
\subfloat{\includegraphics[scale=0.5]{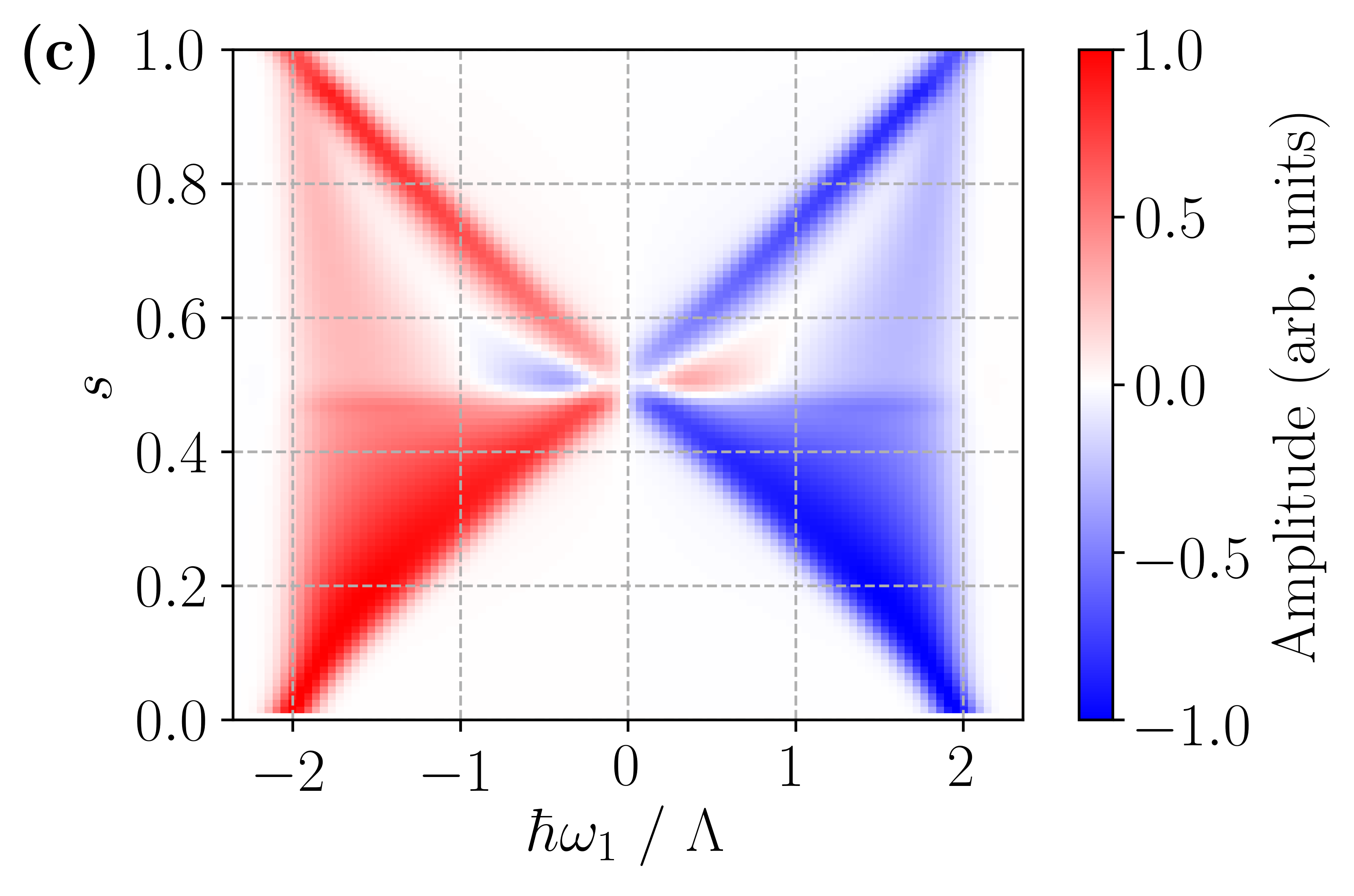}}
\caption{\label{fig:sections_rashba}(a) Diagonal, (b) counterdiagonal and (c) horizontal sections of the imaginary part of the 2D spectra for the lattice Hamiltonian of the Rashba wire $H_\text{Lattice}(w_L, \mu_L, \Delta_L)=H_\text{Lattice}(s \Lambda / 2, (1-s) \Lambda / 2, s \Lambda / 2)$ as a function of $s$ in analogy with Fig.~\ref{fig:sections_ring} in the main text. The dipole operator is given by Eq.~\ref{eq:dipole_lattice} with $\Delta=\Lambda$. For each $s$, the 2D spectra are normalized to their maximal peak amplitude. The chain length is $N=60$. For $s<0.5$, the chain is in the trivial phase, and for $s>0.5$, the chain is in the non-trivial phase. Compared to the results for the pure Kitaev model, we observe only little difference. There is only a slight fading of the high-frequency peaks on the horizontal.}
\end{figure*}

In order to compare these results to the results for the Kitaev chain, we need to fit this continuum model to a lattice model. We arrive at the Kitaev Hamiltonian
\begin{equation}
   H_\text{Lattice} = \sum_{n=1}^N \left[ -w_L a^\dagger_{n+1} a_n - \mu_L a^\dagger_n a_n + \Delta_L a_n a_{n+1}\right] + \text{h.c.}
\end{equation}
with the hopping parameter $w_L$, the chemical potential $\mu_L$ and the $p$-wave pairing term $\Delta_L$ given by
\begin{align}
w_L &= \frac{\hbar^2}{2m\lambda^2}\,  , \\
\mu_L &= \frac{\mu_\text{eff}}{2} - \frac{\hbar^2}{2m\lambda^2} = \frac{\mu}{2} + \frac{g \mu_B \abs{B_z}}{4} - \frac{\hbar^2}{2m\lambda^2}\,  , \\
\Delta_L &= \frac{\Delta_\text{eff}}{\lambda} = \frac{ \hbar u \Delta}{\lambda g \mu_B \abs{B_z}} (e_x + i e_y)\,  .
\end{align}
Here, $\lambda$ is the lattice constant. We introduce the spin-orbit coupling energy $E_\text{SOC} = \hbar u / \lambda$ and the Zeeman energy $E_\text{Zeeman} = g \mu_B \abs{B_z}$. Within this lattice fit the final spectroscopy operator reads
\begin{align}
\label{eq:dipole_lattice}
   \mathbf{d}_\text{Lattice} &= - e \sum_{n} \mathbf{r}_n a^\dagger_{n} a_{n} - e \frac{\abs{\Delta_L}^2}{\abs{\Delta}^2} \sum_{n} \mathbf{r}_n \left( a^\dagger_{n+1} a_n + a^\dagger_n a_{n+1}\right) \\
   &= - e \sum_{n} \mathbf{r}_n a^\dagger_{n} a_{n} - e \frac{E_\text{SOC}^2}{E_\text{Zeeman}^2} \sum_{n} \mathbf{r}_n \left( a^\dagger_{n+1} a_n + a^\dagger_n a_{n+1}\right).
\end{align}
The first term corresponds to the spectroscopy operator in the main text. The additional hopping-like term is suppressed by the ratio of the spin-orbit coupling to the Zeeman energy squared. Equivalently, the prefactor is given by the ratio of the $p$-wave gap in the lattice model to the proximity-induced s-wave gap of the overall wire. In a typical experimental setup this term is at most of the order of a few percent \cite{Kouwenhoven2012}. Such minor terms are  negligible in nonlinear spectroscopy. To demonstrate the influence of this term, we have repeated the calculation of the 2D spectra for the same parameters as used for Fig.~\ref{fig:sections_ring} of the main text. In this setup, the prefactor of the hopping-like contribution varies since we vary $\Delta_L$. We choose $\Delta = \Lambda$. In the topological phase, i.e., $s > 0.5$, the prefactor ranges from $\SI{6.25}{\percent}$ to $\SI{25}{\percent}$. Thus, it is with our intentionally unrealistic choice of $\Delta$ one order of magnitude larger than to be expected. The result is shown in Fig.~\ref{fig:sections_rashba}. We observe no qualitative deviations from the results of the main text. There is merely a fading of the high-frequency peaks along the horizontal. Yet, the peak splitting remains clearly visible.

\end{document}